\documentclass[sigconf]{acmart}

\usepackage{amsmath,amsfonts}
\usepackage{algorithmic}
\usepackage{graphicx}
\usepackage{textcomp}
\usepackage{xcolor}
\usepackage{multirow}
\usepackage{algorithm}
\usepackage{booktabs}
\usepackage{bm}
\usepackage{subfigure}
\AtBeginDocument{%
  \providecommand\BibTeX{{%
    \normalfont B\kern-0.5em{\scshape i\kern-0.25em b}\kern-0.8em\TeX}}}

\copyrightyear{2020} 
\acmYear{2020} 
\setcopyright{acmcopyright}\acmConference[SIGIR '20]{Proceedings of the 43rd International ACM SIGIR Conference on Research and Development in Information Retrieval}{July 25--30, 2020}{Virtual Event, China}
\acmBooktitle{Proceedings of the 43rd International ACM SIGIR Conference on Research and Development in Information Retrieval (SIGIR '20), July 25--30, 2020, Virtual Event, China}
\acmPrice{15.00}
\acmDOI{10.1145/3397271.3401109}
\acmISBN{978-1-4503-8016-4/20/07}

\begin{document}
\fancyhead{}
\title{GAG: Global Attributed Graph Neural Network for Streaming Session-based Recommendation}

\author{Ruihong Qiu}
\affiliation{%
  \institution{The University of Queensland}
  \city{Brisbane}
  \country{Australia}
}
\email{r.qiu@uq.edu.au}

\author{Hongzhi Yin}
\authornote{Corresponding author and contributing equally with the first author.}
\affiliation{%
  \institution{The University of Queensland}
  \city{Brisbane}
  \country{Australia}
}
\email{h.yin1@uq.edu.au}

\author{Zi Huang}
\affiliation{%
  \institution{The University of Queensland}
  \city{Brisbane}
  \country{Australia}
}
\email{huang@itee.uq.edu.au}

\author{Tong Chen}
\affiliation{%
  \institution{The University of Queensland}
  \city{Brisbane}
  \country{Australia}
}
\email{tong.chen@uq.edu.au}

\renewcommand{\shortauthors}{Qiu et al.}

\begin{abstract}
Streaming session-based recommendation (SSR) is a challenging task that requires the recommender system to do the session-based recommendation (SR) in the streaming scenario. In the real-world applications of e-commerce and social media, a sequence of user-item interactions generated within a certain period are grouped as a session, and these sessions consecutively arrive in the form of streams. Most of the recent SR research has focused on the static setting where the training data is first acquired and then used to train a session-based recommender model. They need several epochs of training over the whole dataset, which is infeasible in the streaming setting. Besides, they can hardly well capture long-term user interests because of the neglect or the simple usage of the user information. Although some streaming recommendation strategies have been proposed recently, they are designed for streams of individual interactions rather than streams of sessions. In this paper, we propose a \textbf{G}lobal \textbf{A}ttributed \textbf{G}raph (GAG) neural network model with a Wasserstein reservoir for the SSR problem. On one hand, when a new session arrives, a session graph with a global attribute is constructed based on the current session and its associate user. Thus, the GAG can take both the global attribute and the current session into consideration to learn more comprehensive representations of the session and the user, yielding a better performance in the recommendation. On the other hand, for the adaptation to the streaming session scenario, a Wasserstein reservoir is proposed to help preserve a representative sketch of the historical data. Extensive experiments on two real-world datasets have been conducted to verify the superiority of the GAG model compared with the state-of-the-art methods.
\end{abstract}

\ccsdesc[500]{Information systems~Recommender systems}

\keywords{streaming recommendation, session-based recommendation, graph neural networks}

\maketitle

\section{Introduction}
\label{sec:intro}
\begin{figure}[t]
    \centering
    \includegraphics[width=1.0\linewidth]{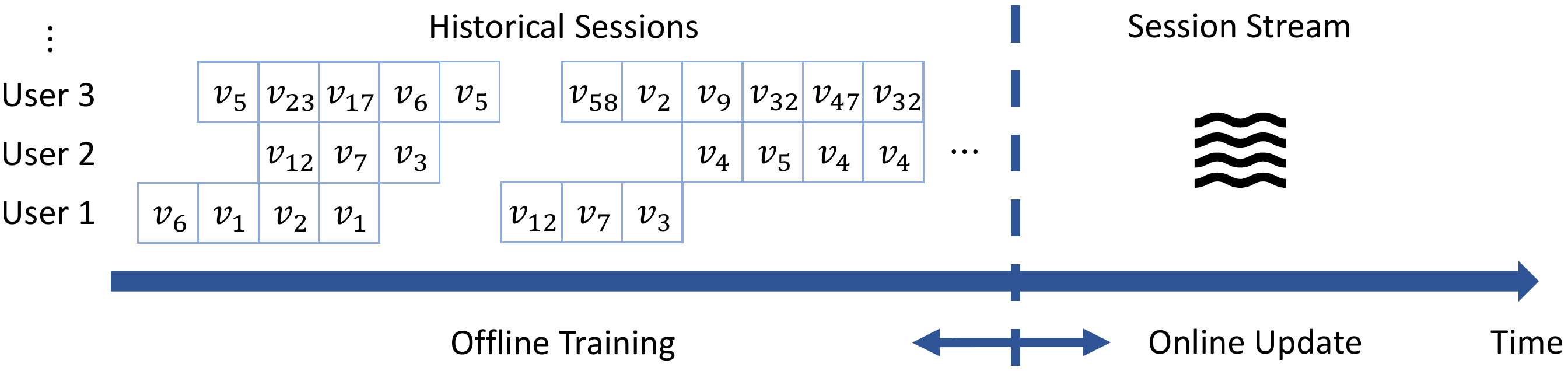}
    \vspace{-0.7cm}
    \caption{The general framework for SSR consists of two phases: the offline training and the online update. During the offline training, the recommender system is trained in a static style with the whole historical session data. When newly generated sessions arrive, the model is expected to conduct an efficient online update with streaming techniques to preserve the users' long-term interests and adapt to the newest preferences.}
    \label{fig:stream}
\end{figure}

In many modern online systems, such as e-commerce and social media platforms, there usually exist a large number of interactions between users and items, such as clicking goods and playing songs. As illustrated in Fig.~\ref{fig:stream}, a sequence of interactions occurring in a certain period can be considered as a session. Session-based recommendation (SR) has been widely studied recently by the academia and the industry~\cite{hidasi2015session,li2017neural,Liu18STAMP,wu2018session,qiu2019rethinking,SunLWPLOJ19}, which aims to recommend items to users based on sessions. However, most of them focus on the static setting, which is not suitable in real-life situation.

In practice, sessions are dynamically produced as a stream, which leads to urgent requirements of streaming session-based recommendation (SSR). As presented in Fig.~\ref{fig:stream}, a general procedure for SSR is to train the recommendation model with the historical sessions to preserve the users' long-term interests and then conduct the online update with the streaming sessions to adapt to their recent preferences. Most of the current research for the SR focuses on the static scenario, where the recommendation models are trained in a batch way. As users' preferences are changing over time, it is infeasible to apply a static model for new coming sessions. To precisely capture the user preference, the model needs to be online updated with the latest sessions. Due to the memory space limit, it is unpractical to stack up the training data by absorbing every new session. In the meanwhile, the training time will be another concern in the streaming scenario, where the recommendation model is expected to be updated promptly. However, a typical SR model needs to train for a long time to converge, which cannot be guaranteed within a short period.

Recently, a few methods utilize the reservoir technique for streaming tasks~\cite{ChenYYC13,Diaz-AvilesDSN12,WangYHLWH18,WangYHWDN18,ChangZTYCHH17}. In these cases, the interaction data is stored with the same probability in the reservoir and then sampled for the online training of the model. However, if streaming sessions are processed in such a manner, the SSR models will suffer from the loss of the session information because the data samples are stored and drawn as discrete interactions. Moreover, the reservoir for traditional streaming tasks is designed to capture the matrix factorization information rather than the session's sequence pattern. Besides, online learning can barely adapt to the session-based recommendation task for newly arrived data as well. For online learning, when a new session comes, the model will update accordingly to capture the recent transition pattern in the latest session. However, these models will easily overfit the new data and fail to maintain users' long-term preferences learned from historical data. Therefore, it is important for the SSR model to effectively exploit the user's information and thus obtain a comprehensive representation for both long- and short-term preferences.

More recently, Guo et al.~\cite{guo2019streaming} applied the reservoir technique to the SSR task with a weighted sampling scheme by evaluating how informative each session is. This method cannot be generalized to other models mainly because it needs to generate an informativeness score for every item in a session with pre-computed item feature vectors, which are commonly unavailable in other models. Moreover, this model directly combines the session-based method with a matrix factorization module for recommendation, which can hardly learn the complicated correlations between users and items in the SSR problem.

To address the issues discussed above, we propose a \textbf{G}lobal \textbf{A}ttributed \textbf{G}raph (GAG) neural network model with a Wasserstein reservoir as a solution to SSR. To make the most usage of the user embeddings and maintain long-term preference information for SSR, we firstly convert a user's session sequence into a session graph having the user embedding as a global attribute associated with the embeddings of interacted items. Based on the global attributed session graph, the GAG model performs graph convolution to learn an updated global attribute, which is passed to the ranking module to output a recommendation list. In the GAG model, the global attribute is applied to effectively assist the joint representation learning of both the entire session and the items within the session. To develop a general reservoir for the SSR problem, we propose the Wasserstein reservoir, which stores and samples session data according to the Wasserstein distance between the generated recommendation lists and the user's real interactions. During the sampling procedure, the Wasserstein reservoir samples the sessions whose recommendation results have a higher Wasserstein distance. Intuitively, the model makes worse predictions in these sessions with a higher Wasserstein distance, which is more informative to refine the model during the online update.

The main contributions of this paper are summarized as follows:
\begin{itemize}
   \item We propose the GAG model to effectively memorize and incorporate users' long-term preferences into the embedding vectors for SSR by treating the user embedding as a global attribute for the session graphs to allow for more expressiveness when learning representations.
  \item A Wasserstein reservoir is designed to actively select the most informative training cases for updating the model in streaming settings. Moreover, our Wasserstein reservoir is an effective yet generic online learning approach that can be easily applied to other steaming session data.
  \item Extensive experimental results on two real-world datasets demonstrate that the proposed GAG model and the Wasserstein reservoir achieve the state-of-the-art performance.
\end{itemize}

\section{Related Work}
\label{sec:rlwork}
\subsection{Session-based Recommendation}
\label{sec:sbr}

{\bf Sequential recommendation} is mainly based on the Markov chain model~\cite{shani2005mdp,zimdars2001using,WangYSCXZ16}, which learns the dependency of items in a sequence data. Using probabilistic decision-tree models, Zimdars et al.~\cite{zimdars2001using} proposed to encode the state of the transition pattern of items. Shani et al.~\cite{shani2005mdp} made use of a Markov Decision Process (MDP) to compute item transition probabilities.

{\bf Deep learning models} are popular with the boom of recurrent neural networks~\cite{hidasi2015session,li2017neural,Liu18STAMP,TangW18,KangM18,MaKL19}. Hidasi et al.~\cite{hidasi2015session} proposed the GRU4REC, which applies the GRU~\cite{chung2014empirical} to treat the data as time series. Some recent approaches use the attention mechanism to avoid the time order. NARM~\cite{li2017neural} stacks GRU as the encoder to extract information and then a self-attention layer to sum up as the session embedding. To further alleviate the bias by time series, STAMP~\cite{Liu18STAMP} replaces the recurrent encoder with the attention layer. Recently, GNN has been widely used in recommendation~\cite{Wang0WFC19,YingHCEHL18,SunZMCGTH19}. Some methods utilize GNN to encode the session information to prevent the misguidance of the session order~\cite{wu2018session,qiu2019rethinking,XuZLSXZFZ19,QiuTOIS20}. SSRM~\cite{guo2019streaming} considers a specific user's historical sessions and applies the attention mechanism to combine them.

\subsection{Streaming Recommendation}
\label{sec:streaming}
{\bf Online learning} focuses on updating the old model with the new data~\cite{HeZKC16,JugovacJK18} to capture the most recent interest of the user~\cite{YinCCHZ15,YinCZWHS16}. For instance, He et al.~\cite{HeZKC16} proposed an element-wise alternative to the least squares technique to address the missing data. Jugovac et al.~\cite{JugovacJK18} applied a replay-based evaluation protocol to update the model with the new arrival events and articles in the news recommendation. Although the models above capture the user's recent interest by updating the model with new interactions, they fail to remember historical interactions.

{\bf Random sampling} is a technique to address the history-ignoring problem by introducing a reservoir to store the user's long-term interactions~\cite{ChenYYC13,Diaz-AvilesDSN12,WangYHLWH18,WangYHWDN18,ChangZTYCHH17,guo2019streaming,Zhang00FX19}. For example, Diaz-Aviles et al.~\cite{Diaz-AvilesDSN12} applied sampling strategies based on active learning principles on the matrix factorization method to update the model. More recently, Guo et al.~\cite{guo2019streaming} used the same reservoir technique to process the streaming sessions.

\subsection{Graph Neural Networks}
\label{sec:gnn}
Originally, GNN is applied basically on directed graphs in a simple situation~\cite{gori2005new,scarselli2009graph}. In recent years, many GNN methods~\cite{kipf2017semi,hamilton2017inductive} work very similar to the message passing network~\cite{gilmer2017neural} to perform an aggregation over the neighborhood of nodes to compute the node embeddings. In~\cite{HamrickABZMTB18,battaglia2018relational,Sanchez-Gonzalez18}, the global attribute is introduced into the GNN layer to maintain a graph level feature in physical systems.

\section{Method}
\label{sec:method}

\subsection{Task Definition}
\label{sec:task}

In the SSR problem, there is an item set $\mathcal{V}=\{v_1,v_2,v_3,\ldots,v_m\}$, where all items are unique and $m$ denotes the number of items. Usually, an embedding layer is applied to represent all items, $\mathbf{x_i}=\text{Embed}_{v}(v_i),i\le m$, where $\text{Embed}_{v}$ is a mapping function that transforms an item into a continuous and dense representation $\mathbf{x_i}\in\mathbb{R}^d$. There is also a user set $\mathcal{U}=\{u_1,u_2,u_3,\ldots,u_n\}$, where all users are unique and $n$ denotes the number of users. Similarly, another embedding layer performs mapping to all user ID, $\mathbf{u_j}=\text{Embed}_{u}(u_j),j\le n$, where $\text{Embed}_{u}$ is another mapping function from a user to $\mathbf{p_j}\in\mathbb{R}^d$. A session sequence at a time step $t$ from a user $u$ is defined as a list $S_{u,t}=[v_1,v_2,\ldots,v_l]$, $v_* \in \mathcal{V}$. $l$ is the length of the session $S$, which may contain duplicated items, $v_a=v_b$, $a, b < l$. In the setting of the SSR, at time step $t$, the recommender system needs to recommend an item $v_{t+1}$ based on $\{S_{u,0},S_{u,1},\ldots,S_{u,t}\}$, which are all sessions of a user from the history to the current. The item $v_{t+1}$ should match the user's preference the most. In the meantime, sessions arrive at a high speed, which means that the computation resource is limited to calculation. As a result, an algorithm should have an efficient way to process the history sessions as well as the current session. Usually, we only recommend the top-$K$ ranked items to users.

\subsection{Overview}
\label{sec:overview}
\begin{figure}[t]
    \centering
    \includegraphics[width=1.0\linewidth]{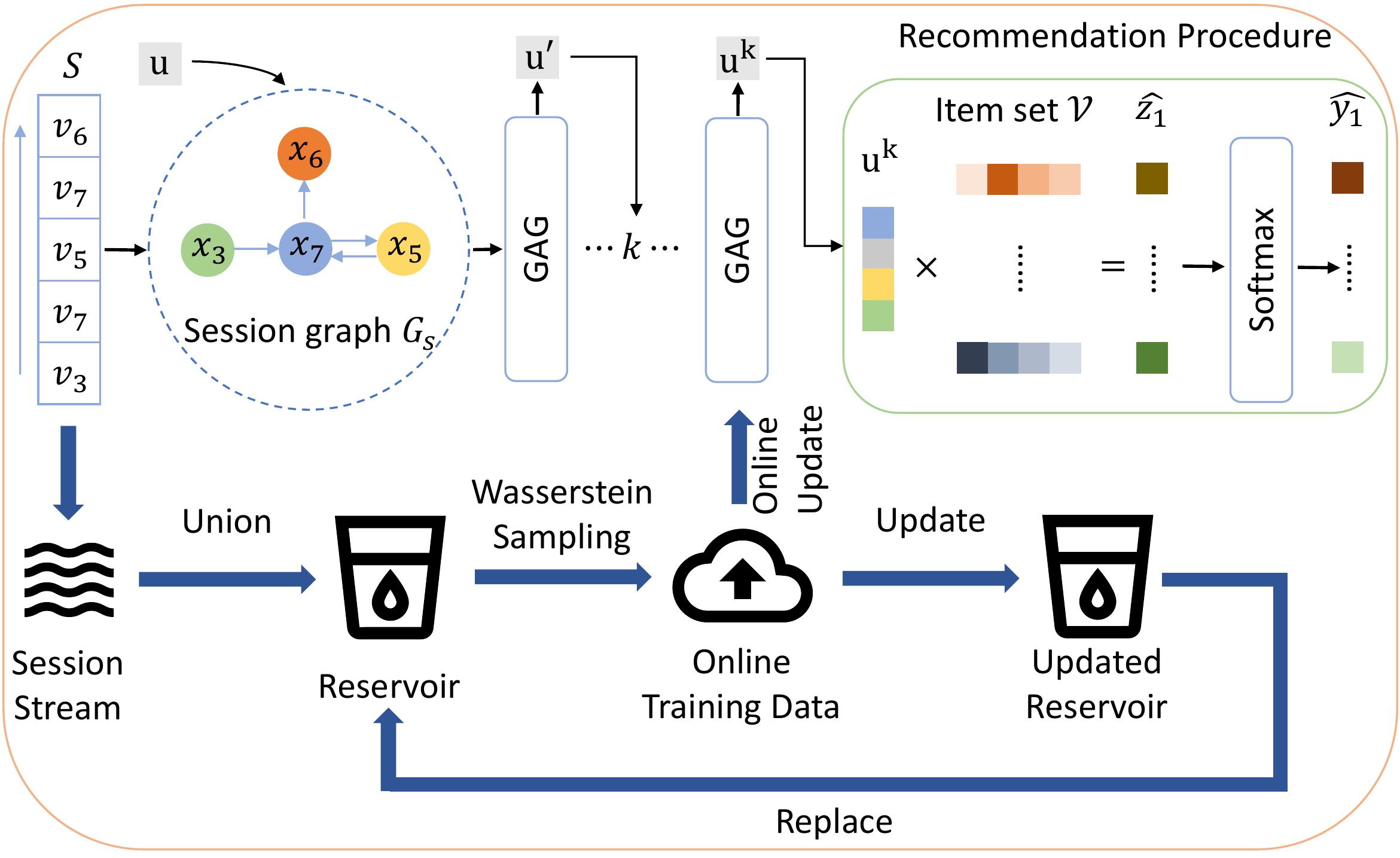}
    \vspace{-0.7cm}
    \caption{The pipeline of the GAG model for the SSR problem. In the upper half, for a specific session $S$ from a user $u$, the GAG model first converts it into a global attributed session graph $G_s$. The GAG layer takes $G_s$ as input and computes the graph convolution based on node features, edge weights and the global attribute. The output of the $k$-layer GAG model is the updated global attribute $\mathbf{u^k}$. To make a personalized recommendation, $\mathbf{u^k}$ is applied to compare with the whole item set to generate the recommendation list. In the bottom half, it shows the procedure of the Wasserstein reservoir dealing with streaming sessions. New streaming sessions and the current reservoir are united together and sampled the online training data according to their respective Wasserstein distance. The online training data is fed to help with the update of the current GAG model. After the online training of the model, the reservoir is updated with the session stream and itself. The updated reservoir is for the following new arrival sessions.}
    \label{fig:whole}
\end{figure}

In this paper, we propose a novel Global Attributed Graph (GAG) neural network model to address the SSR problem mainly by transforming a user's information into the global attribute and incorporating it in the session graph. The architecture of the GAG model is demonstrated in Fig.~\ref{fig:whole}. There are two key components: GAG model for generating recommendation and Wasserstein reservoir for the streaming data learning.

\subsection{Global Attributed Session Graph}
\label{sec:sg}
As shown in Fig.~\ref{fig:whole}, at the first stage, the session sequence is converted into a session graph with a global attribute for the purpose to process each session via GNN. Similar to ~\cite{wu2018session} and~\cite{qiu2019rethinking}, because of the natural order of the session sequence, we convert it into a weighted directed graph. In addition, we incorporate the user's general information as the global attribute $u$ into the session graph, $G_s=(\mathbf{u},V_s,E_s)$, $G_s \in \mathcal{G}$, where $\mathcal{G}$ is the set of all session graphs. In the session graph $G_s$, the node set $V_s$ represents the nodes in the session graph, which are items $v_{s,n}$ from $S$. For every node $\mathbf{v}$, the input feature is the initial embedding vector $\mathbf{x}$. The edge set $E_s$ represents all directed edges $(w_{s,(n-1)n},v_{s,n-1},v_{s,n})$, where $w_{s,(n-1)n}$ is the weight of the edge and $v_{s,n}$ is the click of the item after $v_{s,n-1}$ in $S$. The weight of the edge is defined as the frequency of the occurrence of the edge within the session.

\subsection{Global Attributed Graph Neural Network}
\label{sec:gag}
With the construction of the global attributed session graph $G_s$, we propose the GAG model to perform graph convolution on $G_s$ with the node features, edge features and the global attribute. When the GAG model is fed with the session graph as the input, the computation proceeds from the edge, the node to the global attribute.

\begin{figure}[t]
    \centering
    \includegraphics[width=1.0\linewidth]{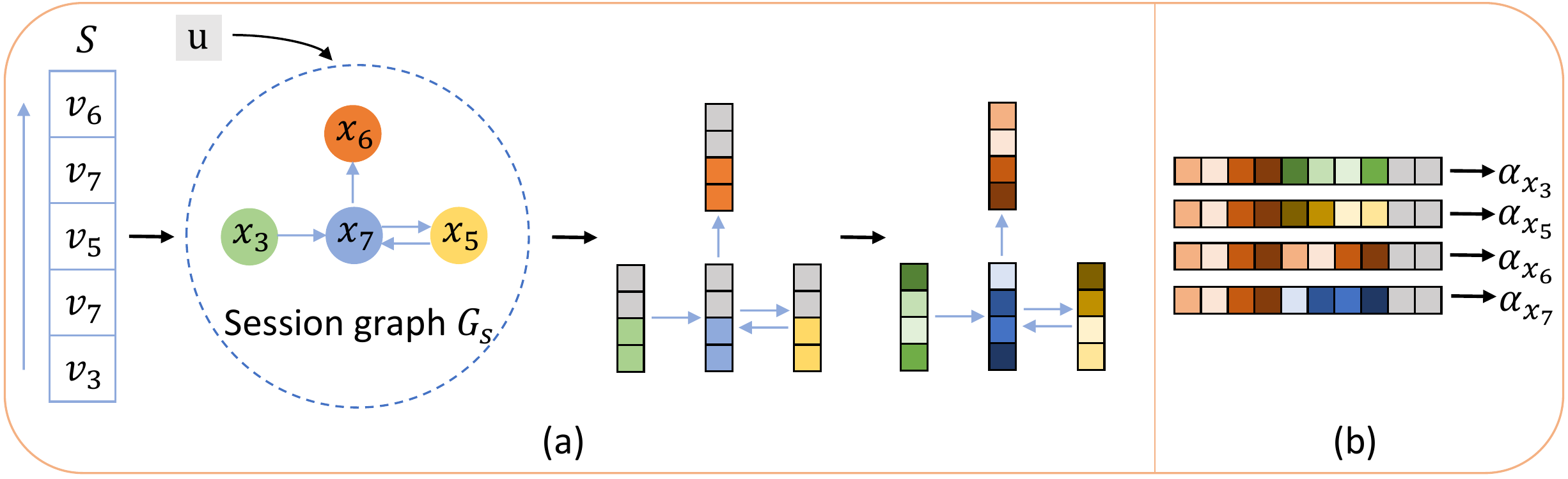}
    \vspace{-0.7cm}
    \caption{The usage of the global attribute in GAG. (a) In the input stage, the global attribute is concatenated with the node feature for every node, which gives out a node feature concatenated with the global attribute. (b) In the global attribute update procedure, the attention weight $\alpha _i$ is calculated based on the concatenation of the features of the last node $\mathbf{v_l}$, the individual node $\mathbf{v_i}$ and the global attribute $\mathbf{u}$ itself.}
    \label{fig:gag}
\end{figure}

First, the per-edge update is calculated among all edges to compute the output features from sender nodes $\mathbf{v_{s_k}}$ to receiver nodes $\mathbf{v_{r_k}}$ with additional features of the edge itself $\mathbf{e_k}$ and the global attribute $\mathbf{u}$.
In our design of GAG, the edge feature, i.e., the weight of the edge, will not be updated because the edge feature is not in the dense vector form. This setting means that the output of the edge update function
$\mathbf{e_{k}}^{\prime}$, will only be used to update other node features and global features.
The output new session graph $G_s^{\prime}$ has the same edge set $E_s$ as the input session graph $G_s$. Because the session graph is built in the directed situation, we compute the propagation in both directions to represent the different meanings for a node as a sender and a receiver in an edge. Therefore, the $\phi^{e}$ function is designed as:
\begin{equation}
\setlength{\belowdisplayskip}{0pt}
\label{eq:ekin}
\begin{aligned}
    \mathbf{e_{k,in}}^{\prime}&=\phi_\text{in}^{e}\left(\mathbf{e_{k}}, \mathbf{v_{r_{k}}}, \mathbf{v_{s_{k}}}, \mathbf{u}\right)\\
    &=w_k\cdot\text{MLP}(\mathbf{v_{s_{k}}}||\mathbf{u}),
\end{aligned}
\end{equation}
\begin{equation}
\setlength{\abovedisplayskip}{0.cm}
\setlength{\belowdisplayskip}{3pt}
\label{eq:ekout}
\begin{aligned}
    \mathbf{e_{k,out}}^{\prime}&=\phi_\text{out}^{e}\left(\mathbf{e_{k}}, \mathbf{v_{r_{k}}}, \mathbf{v_{s_{k}}}, \mathbf{u}\right)\\
    &=w_k\cdot\text{MLP}(\mathbf{v_{s_{k}}}||\mathbf{u}),
\end{aligned}
\end{equation}
where $w_k$ is the scalar form of $\mathbf{e_{k}}$, MLP stands for the multi-layer perceptron to encode the features provided by a concatenation of the sender and receiver node and $||$ means the concatenation between two vectors along the unit dimension. MLPs in both equations are not shared because they perform different operations to the node features. In Fig.~\ref{fig:gag} (a), the procedure of exploiting the global attribute in the node update function is demonstrated in detail.

After updating the new per-edge features, per-node features are updated based on the per-edge features when the node is the sender and the receiver. The new per-node feature consists of the normalized summation of the in-coming and out-going neighborhoods. The aggregation procedures are as:
\begin{equation}
\label{eq:innode}
    \mathbf{v_{i,in}}^{\prime}=\sum_{j\in\{\mathbf{v_{s_{j}}}=\mathbf{v_{r_{i}}}\}} \frac{\mathbf{e_{j,out}}^{\prime}}{\sqrt{N_{in}(i)N_{out}(j)}},
\end{equation}
\begin{equation}
\label{eq:outnode}
    \mathbf{v_{i,out}}^{\prime}=\sum_{j\in\{\mathbf{v_{r_{j}}}=\mathbf{v_{s_{i}}}\}} \frac{\mathbf{e_{j,in}}^{\prime}}{\sqrt{N_{out}(i)N_{in}(j)}},
\end{equation}
where $N_{in}(*)$ and $N_{out}(*)$ represent the in-coming and the out-going degree of a node.

The final result of the neighborhood aggregation is a linear transformation of the in-coming and the out-going feature:
\begin{equation}
\label{eq:node}
    \mathbf{v_i}^{\prime}=\text{MLP}(\mathbf{v_{i,in}}^{\prime}||\mathbf{v_{i,out}}^{\prime}).
\end{equation}
The updated node feature $\mathbf{v_{i}}^{\prime}$ actually includes the information from the node feature of itself and the neighborhood, the edge weight and the global attribute.

At the last step of the GAG layer forward computations, the global attribute is updated based on all the features of nodes, edges and the global attribute itself in the graph. It is worth noting that the purpose of the session-based recommendation is to generate a representation of a session to recommend items. Therefore, the final global attribute is exactly the representation we desire to represent the whole graph. Similar to the previous methods to separate the representation of the preference in long-term and short-term parts inside a session~\cite{li2017neural,Liu18STAMP,wu2018session}, a self-attention on the last input item $v_l$, of the session is applied to aggregate all item features of the session to be the session-level feature. The computation of updating $\mathbf{u}$ to $\mathbf{u_{sg}}$ is defined as:
\begin{equation}
    \begin{aligned}
    \mathbf{u}^{\prime}&=\phi^{u}\left(V^{\prime}, \mathbf{u}\right)\\
    &=\text{Self-Att}(v_l^{\prime},v_i^{\prime},\mathbf{u})+\mathbf{u},
    \end{aligned}
\end{equation}
where $v_i\in V^\prime,i=1,2,3,\ldots,l$ represent all items in the session after being updated to the new features. In the setting of the session graph, items are converted into nodes and the Self-Att can be divided into two steps:
\begin{equation}
\label{eq:att_weight}
    \alpha_{i}=\text{MLP}(\mathbf{v_{l}}^{\prime}||\mathbf{v_{i}}^{\prime}||\mathbf{u}),
\end{equation}
\begin{equation}
\label{eq:usg}
    \mathbf{u_{sg}}=\sum_{i=1}^{n} \alpha_{i} \mathbf{v_{i}}^{\prime},
\end{equation}
where an MLP is utilized to learn the weights that aggregate the node features, the last node feature and the global attribute. In Fig.~\ref{fig:gag} (b), the detail of the computation of attention weights is presented.

Besides, because the user's profile is applicable in the SSR setting, the incorporation of the user embedding can provide the extra user information. Therefore, the final formula for how to compute the output of the global attribute with user information is defined as:
\begin{equation}
\label{eq:u_prime}
    \mathbf{u}^{\prime}=\mathbf{u_{sg}}+\mathbf{u}.
\end{equation}
The residual addition can help to alleviate the burden of directly learning the updated global attribute.

\subsection{Recommendation}
\label{sec:rec}
The last stage of the GAG to perform the recommendation is the generation of candidate items based on the representation of the input session and the user's profile.
We compute a score for every item and form a score vector $\hat{\mathbf{z}}\in\mathbb{R}^n$, where $\text{n}$ is the size of the item set. Specifically, the score vector $\hat{\mathbf{z}}$ is calculated as:
\begin{equation}
\label{eq:rec}
    \hat{\mathbf{z}}=\mathbf{u^\prime}^\top \mathbf{X},
\end{equation}
where $\mathbf{X}$ is embeddings of all items in the item set.

The probabilistic form of the prediction $\hat{\mathbf{y}}$ is defined as:
\begin{equation}
    \hat{\mathbf{y}}=\text{Softmax}(\hat{\mathbf{z}}).
\end{equation}

\subsection{Wasserstein Reservoir for Streaming Model}
\label{sec:res}
In this section, we extend our offline model to the streaming setting. Our purpose is to update our model with the new arrival session data while keeping the knowledge learned from the historical sessions. Traditionally, online learning methods update the model only with the new data, which will always lead to forgetting the past~\cite{RendleS08}. To prevent the model from losing the awareness of historical data, we leverage the reservoir to maintain a long-term memory of the historical data~\cite{guo2019streaming,WangYHLWH18,WangYHWDN18,ChenYYC13}. The reservoir technique is widely used in the streaming database management systems.

The purpose of applying a reservoir is to maintain a representative sketch of all the historical data. Therefore, we conduct a random sampling~\cite{Vitter85Random} to select the data stored in the reservoir. Let $C$ denote the reservoir, which contains $|C|$ sessions. Let $t$ be the time order of the arrival session instance. When $t>|C|$, the reservoir will store this $t$-th session with the probability:
\begin{equation}
\label{eq:store}
    p_\text{store}=\frac{|C|}{t},
\end{equation}
and replaces a uniformly random session that is already in $C$. This method of generating the reservoir is actually sampling randomly from the current dataset, and it can successfully maintain the model’s long-term memory~\cite{Diaz-AvilesDSN12}.

Although the reservoir can be updated in the way introduced above, the probability for a new arrival session to be included tends to be smaller over time, and the reservoir will have a chance of overlooking the recent data. However, the recent data is crucial for predicting the user's varying preference. Besides, new users and new items are exposed to the system continually. In consequence, when the new session data $C^{new}$ arrives, we update the pre-trained model with $C^{new}$ and the reservoir $C$.

The reservoir sampling procedure above enables the model to continually update according to the new and old data. However, it can narrowly achieve a good performance in reality. The reason is that most training sessions in $C$ are already learned well by $M$, which results in $C^{rand}$ mainly containing helpless training samples. Actually, if the current model makes a worse prediction on a session, it is more worthwhile to update the model with this session because it either contains the latest preference of a user or there is some item transition patterns that the current model cannot learn well. Such a session is called an \textit{informative} session to the model and this session is more significant to the model update. In our work, the \textit{informativeness} of a session is defined as the distribution distance $d$ between its predicted recommendation result $\hat{\mathbf{y}}$ and the real interaction $\mathbf{y}$. $\hat{\mathbf{y}}$ is a distribution given by the model $M$ and $\mathbf{y}$ is a one-hot vector. Intuitively, the greater of this distance, $M$ predicts the worse over the session. There are different distance metrics of two distributions:
\begin{itemize}
    \item The \textit{Wasserstein} distance (EMD distance)~\cite{RubnerTG00}:
    \begin{equation}
        d_W(\mathbb{P}_{r}, \mathbb{P}_{g})=\inf _{\gamma \in \Pi\left(\mathbb{P}_{r}, \mathbb{P}_{g}\right)} \mathbb{E}_{(x, y) \sim \gamma}[\|x-y\|]
    \end{equation}
    \item The \textit{Kullback-Leibler} (KL) divergence~\cite{kullback1951}:
    \begin{equation}
        d_{K L}(\mathbb{P}_{r}\|\mathbb{P}_{g})=\sum_{i=1}^{n} P_r(x) \log \frac{P_r(x)}{P_g(x)}
    \end{equation}
    \item The \textit{Total Variation} (TV) distance:
    \begin{equation}
        d_{T V}(\mathbb{P}_{r}, \mathbb{P}_{g})=\sup _{A \in \Sigma}\left|\mathbb{P}_{r}(A)-\mathbb{P}_{g}(A)\right|
    \end{equation}
\end{itemize}

Within the recommendation task, the real distribution $P_r$ is always a one-hot vector $\mathbf{y}$. Under this situation, the KL divergence between $\mathbf{y}$ and $\hat{\mathbf{y}}$, $d_{K L}(\mathbf{y}\|\hat{\mathbf{y}})$ is:
\begin{equation}
    d_{K L}(\mathbf{y}\|\hat{\mathbf{y}})=- \log P_g(v_i),
\end{equation}
where $P_g(v_i)$ is the predicted probability over the ground truth item $v_i$ given by the model $M$. Therefore, the KL divergence fails to take the whole distribution into consideration.

As for the TV distance, $d_{T V}(\mathbf{y}, \hat{\mathbf{y}})$ is:
\begin{equation}
    d_{T V}(\mathbf{y}, \hat{\mathbf{y}})=\max_{j \neq i}(1-P_g(v_i),P_g(v_j)),
\end{equation}
where $P_g(v_j)$ is the predicted probability over the item $v_j$ given by the model $M$ other than the ground truth item $v_i$. The TV distance either captures the difference over the real interacted item or other unrelated items.

Obviously, both KL divergence and TV distance have drawbacks of focusing on a certain elementary event while neglecting the whole distribution. The only metric that can preserve the difference at each prediction score is the Wasserstein distance.

Therefore, we propose a Wasserstein reservoir construction strategy that samples the session whose output probability distribution over the recommendation item has a higher Wasserstein distance to the user's real interaction with a higher probability. Intuitively, when the output probability of a session has higher Wasserstein distance, the output will contain more information compared with those with lower Wasserstein distance. Therefore, we sample sessions according to the Wasserstein distance of their output probabilities. For session $s_i$ with corresponding Wasserstein distance $d_i$ for its output, the sampling probability is calculated as follows:
\begin{equation}
\label{eq:prob}
    p_\text{sample}(s_i)=\frac{d_i}{\sum_{s_j\in C\cup C^{new}-S} d_j},
\end{equation}
where $S$ is the ongoing updated dataset. The detailed construction of the Wasserstein reservoir is shown in Algorithm~\ref{alg:res}. During the sampling procedure, because there are always new items and new users in the streaming data, their corresponding embedding vectors are not trained before they show up. To prevent the model from neglecting the new sessions, these sessions are directly included.
\begin{algorithm}[t]
    \caption{Online Update with Wasserstein Reservoir}
    \label{alg:res}
    \begin{algorithmic}[1]
    \REQUIRE the current time step $t$, the current model $M$, the current reservoir $C$ and new sessions $C^{new}$;\
    \ENSURE the updated model $M^{\prime}$ and the updated reservoir $C^{\prime}$;\
    \STATE initialize a blank update dataset $S$;
    \IF{the last epoch finished}
    \FOR{each session $s_i$ in $C^{new}$}
    \IF{a new item or a new user appears}
    \STATE append $s_i$ to $S$;
    \ENDIF
    \ENDFOR
    \FOR{each session $s_i$ in $C\cup C^{new}-S$}
    \STATE compute the Wasserstein distance $d_i$ of $s_i$;
    \ENDFOR
    \STATE compute the sample probability $p(s_i)$;
    \STATE sample the the rest of $S$ according to Eq.~\eqref{eq:prob};
    \ENDIF
    \STATE update the current model $M$ with $S$ to $M^{\prime}$;
    \FOR{each session $s_i$ in $C^{new}$}
    \STATE update the reservoir with $s_i$ according to Eq~\eqref{eq:store} to $C^{\prime}$;
    \STATE update $t$;
    \ENDFOR
    \end{algorithmic}
\end{algorithm}

\subsection{Training}
\label{sec:train}
Since the recommendation task is considered as a classification problem over the whole item set, we can apply a multi-class cross-entropy loss between the predicted recommendation distribution $\hat{\mathbf{y}}$ and the real interaction $\mathbf{y}$:
\begin{equation}
    L=-\sum_{i=1}^{l}\mathbf{y}_i\log \left(\hat{\mathbf{y}}_i\right),
\end{equation}
where $l$ is the number of training sessions in a mini-batch.

\section{Experiment setup}
\label{sec:exp-setup}

\subsection{Dataset}
\label{sec:dataset}
\textit{LastFM} \footnote{http://mtg.upf.edu/static/datasets/last.fm/lastfm-dataset-1K.tar.gz} is a real-world music recommendation dataset, which is released by Celma Herrada~\cite{Celma10Music}. In this work, we mainly focus on music artist recommendation. As shown by Guo et al.~\cite{guo2019streaming}, we also consider doing recommendations on artists and choose the 10,000 most popular ones. Based on the time order, we group transactions in 8 hours from the same user as a session. Following~\cite{li2017neural}, sessions that contain more than 20 transactions or less than 2 will be filtered out. In total, there are 298,919 sessions after the pre-processing.

\textit{Gowalla} \footnote{https://snap.stanford.edu/data/loc-gowalla.html} is a point-of-interest real-world dataset collected from a social network for users' check-in. The same as Guo et al.~\cite{guo2019streaming}, the 30,000 most popular places are used for experiments and check-ins within 1 day are defined as a session. Again, sessions that contain more than 20 transactions or less than 2 will be filtered out. Finally, we have 198,680 sessions during experiments.

\subsection{Metrics}
\label{sec:metrics}
Following Chang et al.~\cite{ChangZTYCHH17} and Guo et al.~\cite{guo2019streaming}, to simulate the streaming situation of the data arriving situation, the dataset is split into two proportions (60\% and 40\%) by the chronological order of all data. The first part is defined as the training set ($\mathcal{D}^{train}$) while the second part is the candidate set ($\mathcal{D}^{candidate}$). Specifically, $\mathcal{D}^{train}$ is used for training the GAG model as offline data. As for $\mathcal{D}^{candidate}$, it is designed to simulate the online streaming session data. Especially for $\mathcal{D}^{candidate}$, it is further divided into five same-sized test set by time order, $\mathcal{D}^{test,1},\ldots,\mathcal{D}^{test,5}$. Test sets are provided in the time order to the model during test time. And after testing on the current test set, the model will be updated according to it and the updated model accounts for the test for the next test set. Such an online update is designed for the streaming occasion.

To evaluate the performance of our model, according to the nature of the user's picking of the first few recommended items, the top-20 recommendation is applied here and we mainly compare different models based on the \textbf{Recall@$K$} and \textbf{MRR@$K$}.

\subsection{Baselines}
\label{sec:bsl}
In our experiments, we will mainly compare our GAG model to the following representative baseline methods:
\begin{itemize}
   \item \textbf{POP} always chooses the most popular items of all users to recommend to other users. It is a simple yet strong baseline.
   \item \textbf{S-POP} recommends the most popular items that appear in the current session instead of the whole item set.
   \item \textbf{BPR-MF} is a method that mainly makes use of a pairwise ranking loss~\cite{Rendle09Bayesian}. Also, the Matrix Factorization is modified to suit the session-based recommendation as in~\cite{li2017neural}.
   \item \textbf{GRU4REC} utilizes GRU layers to learn the session embedding in the anonymous setting~\cite{hidasi2015session}.
   \item \textbf{NARM} adds an attention layer to item level across the session to encode the session information within an anonymous setting~\cite{li2017neural}.
   \item \textbf{FGNN} makes use of GNN to learn an embedding of an anonymous session to make recommendation~\cite{qiu2019rethinking}. This method does not consider the user information.
   \item \textbf{SSRM} is a state-of-the-art method for the SSR problem, which applies a reservoir to sample the history sessions to help the current session embedding learning~\cite{guo2019streaming}.
\end{itemize}

\subsection{Training Detail}
\label{sec:detail}
In the implementation of the model, we set all MLPs with 1 layer and the embedding size is 200 for the fairness of comparison. We use Adam~\cite{adam} with a learning rate of 0.003 and set the batch size as 100 to train the GAG model. The size of the reservoir is set to $|D|/100$ and the window size is set to $|C^{new}|/2$ on each $C\cup C^{new}$.

\section{Experiment Results}
\label{sec:exp}
In this section, we will describe our experiments on two real-world datasets and demonstrate the efficacy of our proposed model GAG. Specifically, four research questions will be addressed:
\begin{itemize}
   \item \textbf{RQ1} How does our proposed GAG model perform compared with current state-of-the-art methods? (Section~\ref{sec:rq1})
   \item \textbf{RQ2} How does the global attribute help to solve the SSR problem? (Section~\ref{sec:rq2})
   \item \textbf{RQ3} How is the performance of the Wasserstein reservoir? (Section~\ref{sec:rq3})
   \item \textbf{RQ4} How is the parameter sensitivity of the GAG model? (Section~\ref{sec:rq4})
\end{itemize}

\subsection{Comparisons with Baseline Methods}
\label{sec:rq1}
\begin{figure}[t]
    \centering
    \subfigure[Recall@20 on Gowalla.]{
    \label{fig:r20-gowalla}
    \includegraphics[width=0.46\linewidth]{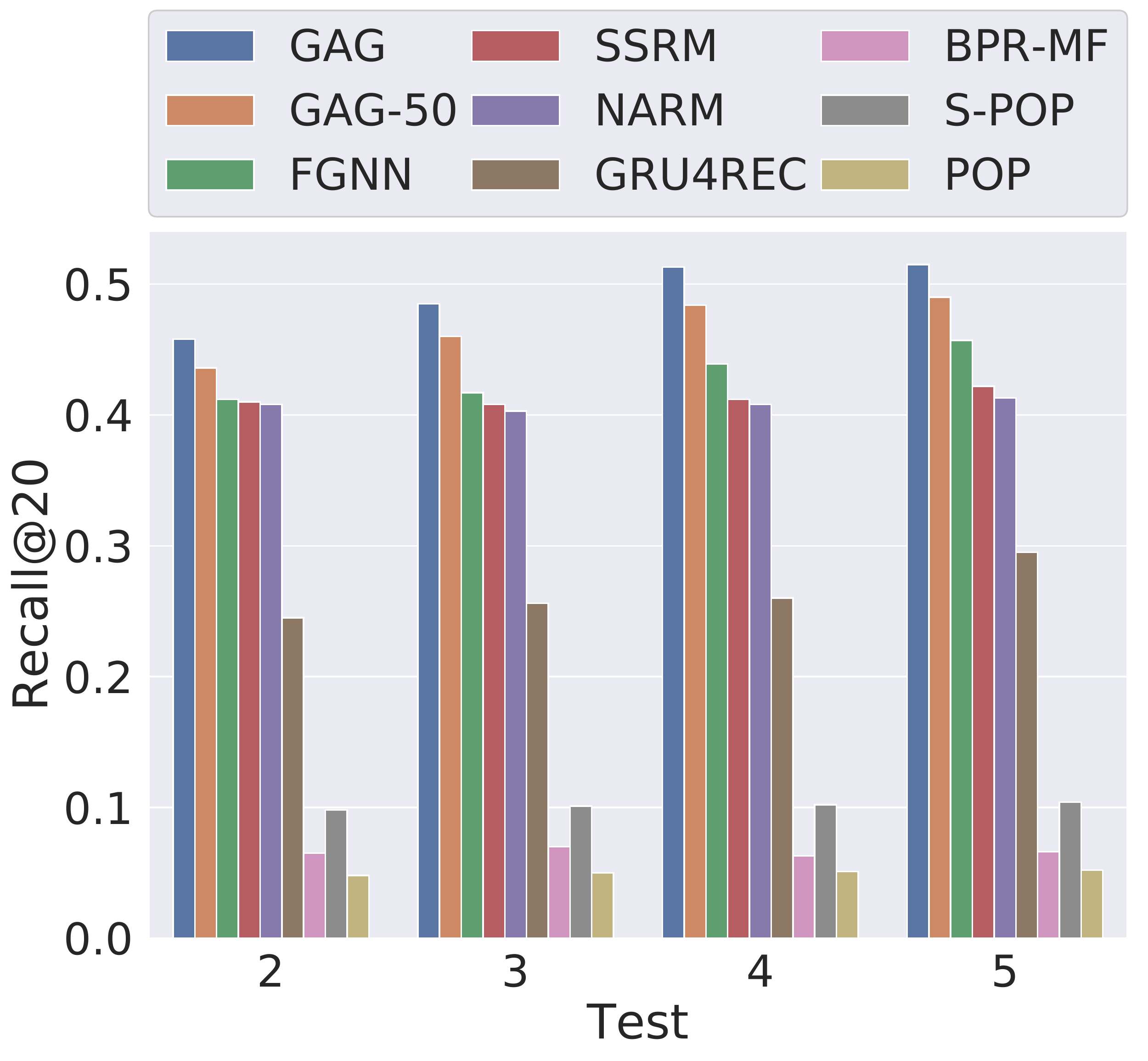}
    }
    \subfigure[MRR@20 on Gowalla.]{
    \label{fig:mrr20-gowalla}
    \includegraphics[width=0.47\linewidth]{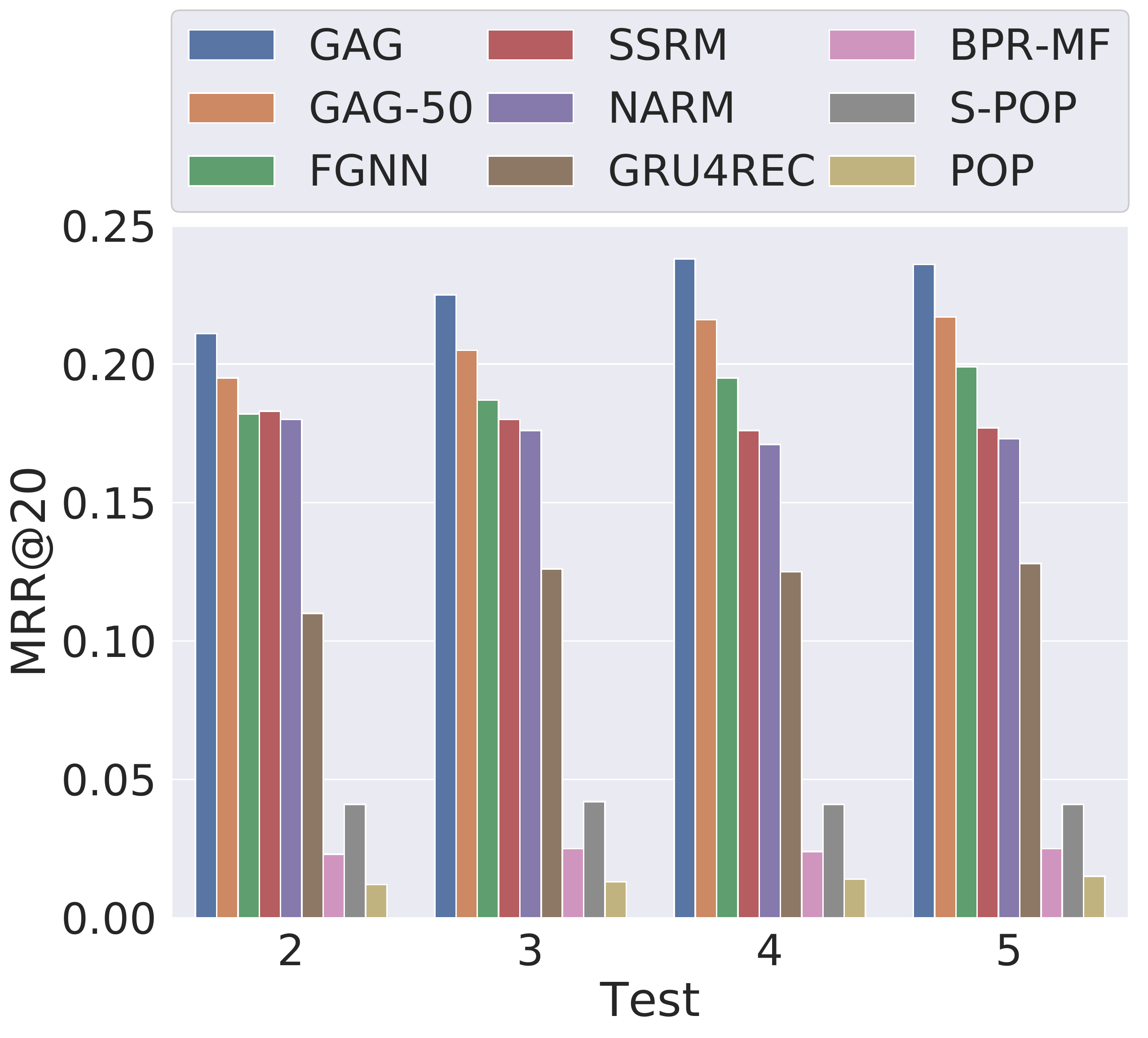}
    }
    \subfigure[Recall@20 on LastFM.]{
    \label{fig:r20-lastfm}
    \includegraphics[width=0.465\linewidth]{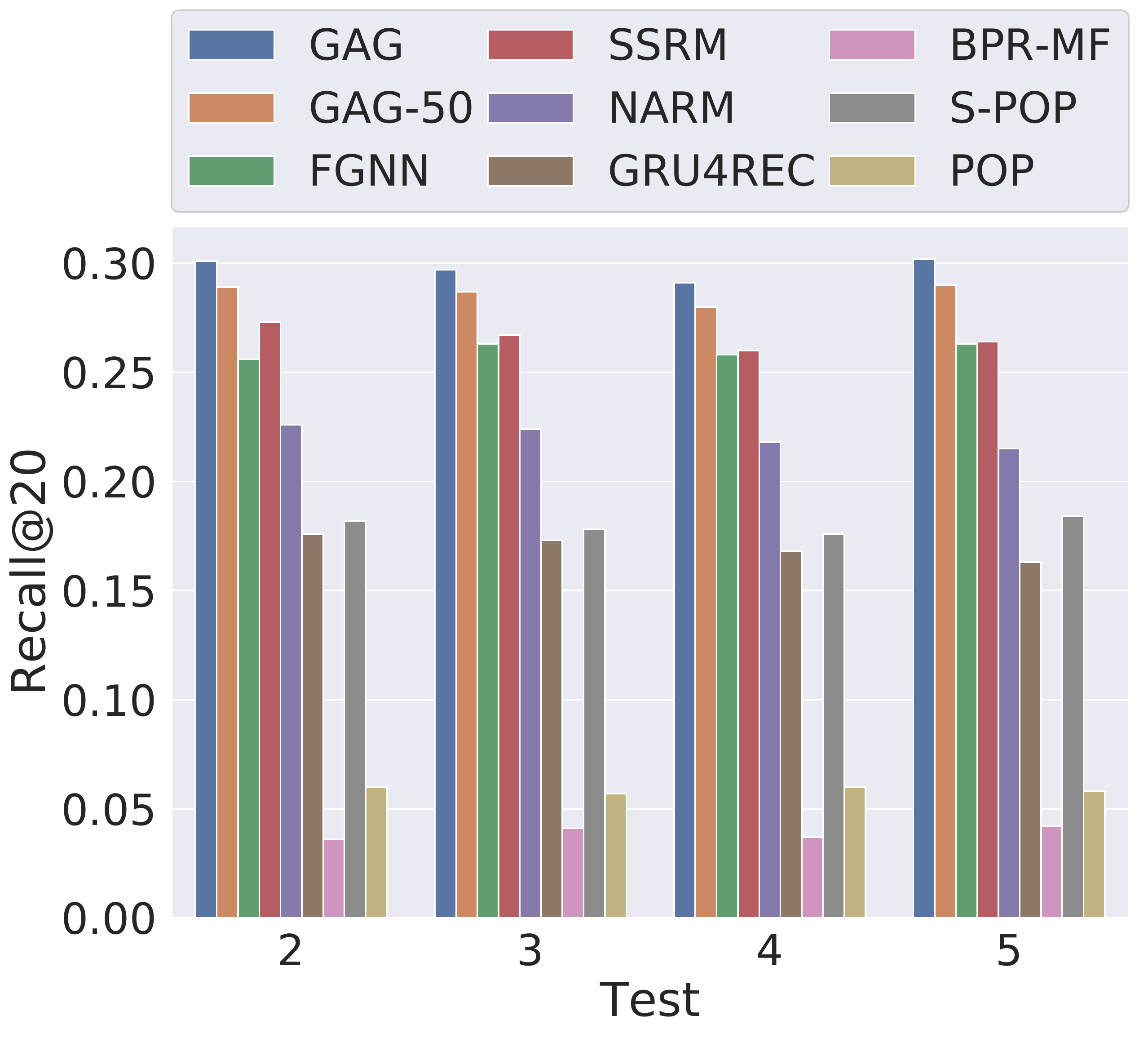}
    }
    \subfigure[MRR@20 on LastFM.]{
    \label{fig:mrr20-lastfm}
    \includegraphics[width=0.465\linewidth]{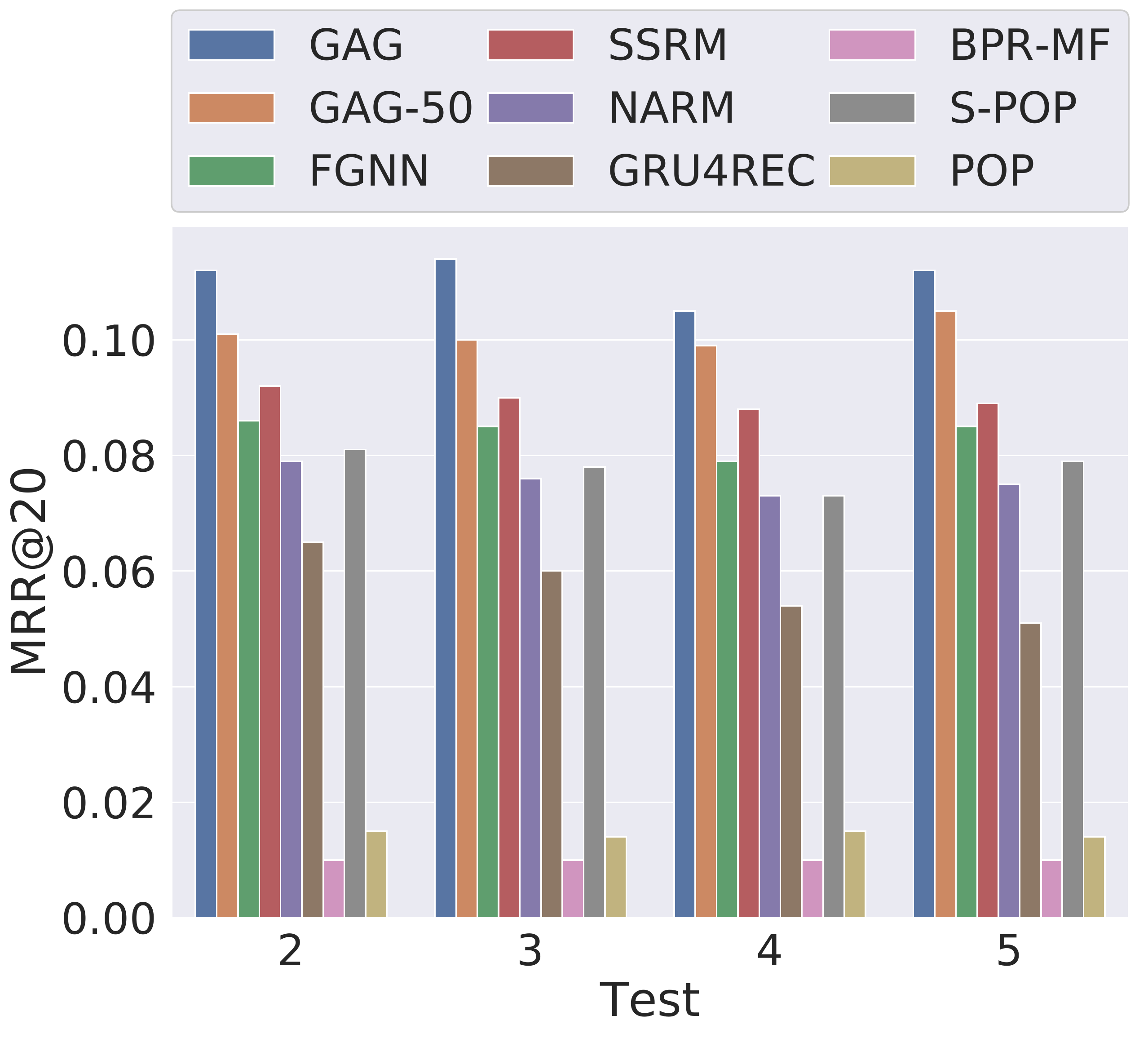}
    }
    \vspace{-0.5cm}
    \caption{Results of streaming session-based recommendation performance.}
    \label{fig:gen}
\end{figure}

To evaluate the overall performance of the GAG, we compare the GAG model with the baseline methods mentioned in Section~\ref{sec:bsl} by the Recall@20 and MRR@20 scores on Gowalla and LastFM datasets. The overall results are demonstrated in Fig.~\ref{fig:gen}. We also use the top-5 and top-10 recommendation results for a more in-depth comparison. For fairness, we have the GAG-50 model with the embedding size as 50 in accordance with the baseline methods. For the state-of-the-art performance, we have the GAG model with the embedding size as 200.

\subsubsection{\textbf{General Comparison}}
The overall performance is shown in Fig.~\ref{fig:gen}. Clearly, the proposed GAG-50 model outperforms all the baseline methods in all situations. With the embedding size increasing to 200, the GAG model achieves the state-of-the-art results. Both of them show the superiority of the GAG model.

The performance is much worse for conventional methods, such as POP and S-POP, both of which recommend the most popular items to users. POP recommends the most popular ones from the whole item set while S-POP chooses the most popular item in the current session. POP fixes the recommendation list, which fails to detect the different patterns of users' behaviors in different sessions. However, S-POP is still a strong session-based baseline method because it can capture the item's re-occurrence patterns of the session. Besides, for the shallow method BPR-MF, which performs a matrix factorization of the whole user-item interaction matrix, it has higher performance compared with POP because it can perform the personalized recommendation. However, BPR-MF still fails to outperform S-POP in the SSR problem because S-POP can further extract the session-specific information.

For deep learning models, GRU4REC is a method that utilizes GRU to process the session as a sequence and output a session embedding to make a recommendation. It outperforms traditional methods in most situations, which is proof of the superiority of the deep learning-based approaches. Besides, methods utilizing the attention mechanism, e.g., NARM and SSRM, obtain a great improvement compared with GRU4REC, which shows the capability of the attention mechanism to learn the inter-dependency of items. Especially, SSRM is specifically for the SSR problem and it is the strongest baseline in the experiment.

\begin{figure}[t]
    \centering
    \subfigure[Recall@5 index.]{
    \label{fig:r5-gowalla}
    \includegraphics[width=0.46\linewidth]{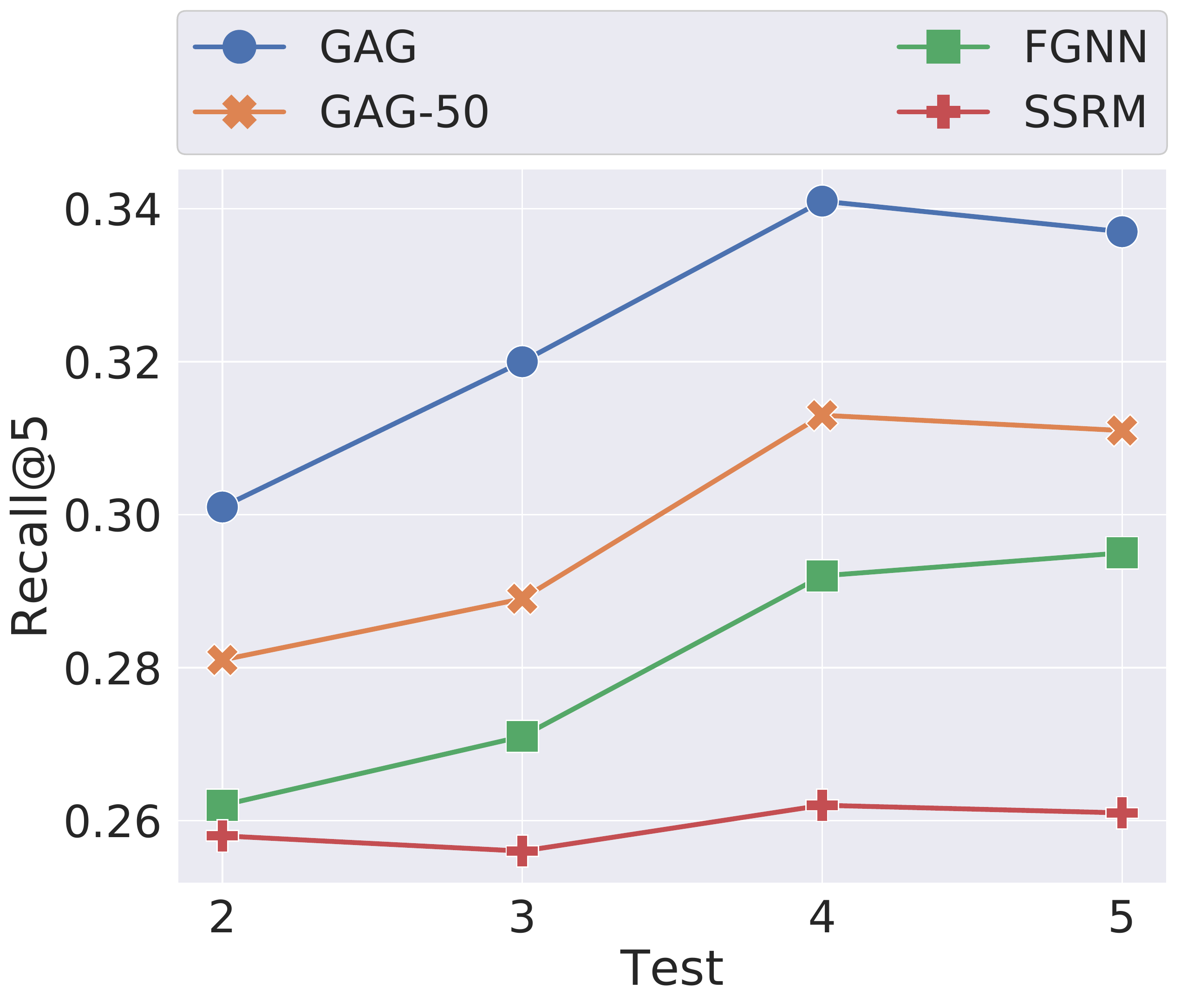}
    }
    \subfigure[MRR@5 index.]{
    \label{fig:mrr5-gowalla}
    \includegraphics[width=0.47\linewidth]{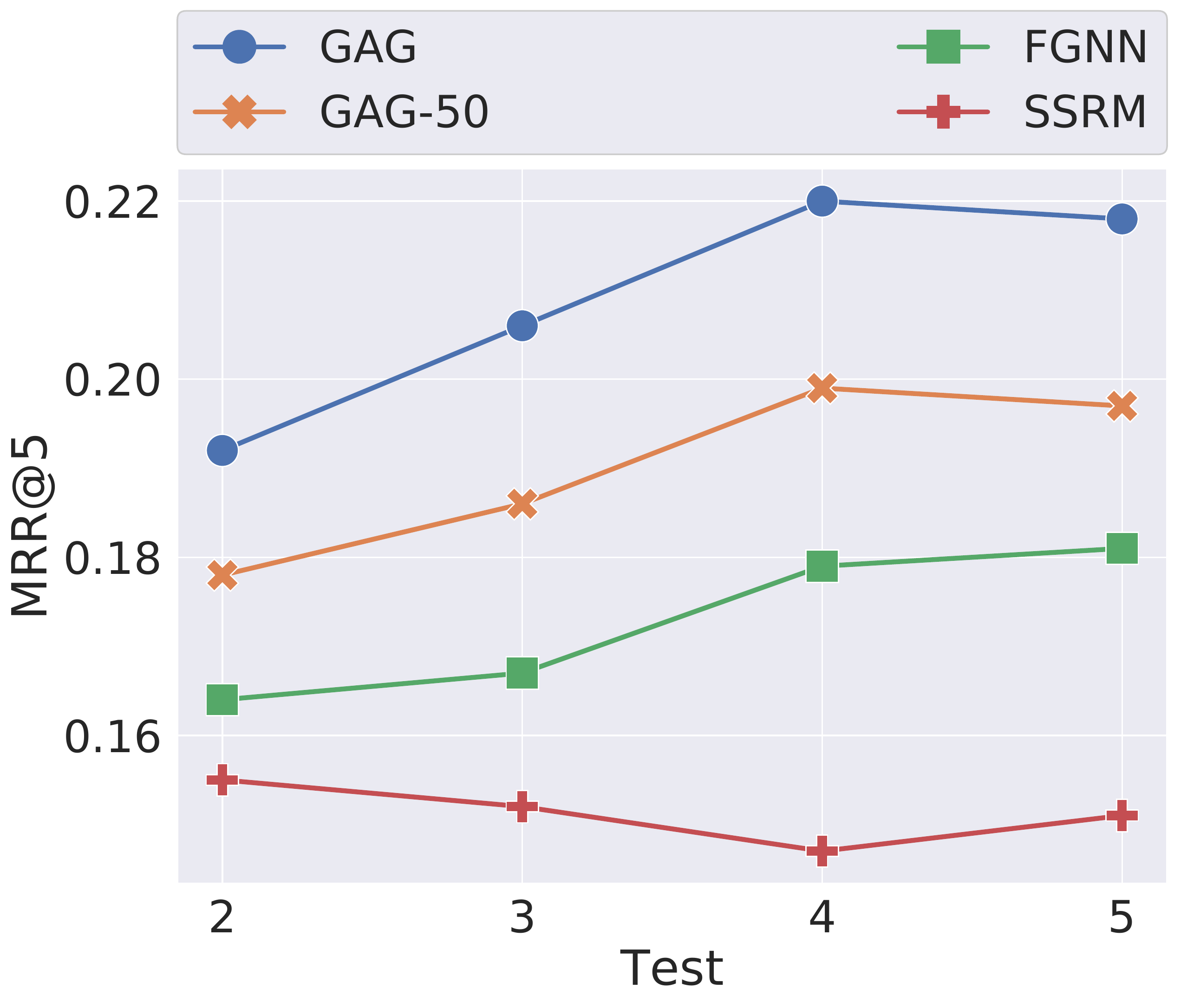}
    }
    \subfigure[Recall@10 index.]{
    \label{fig:r10-gowalla}
    \includegraphics[width=0.46\linewidth]{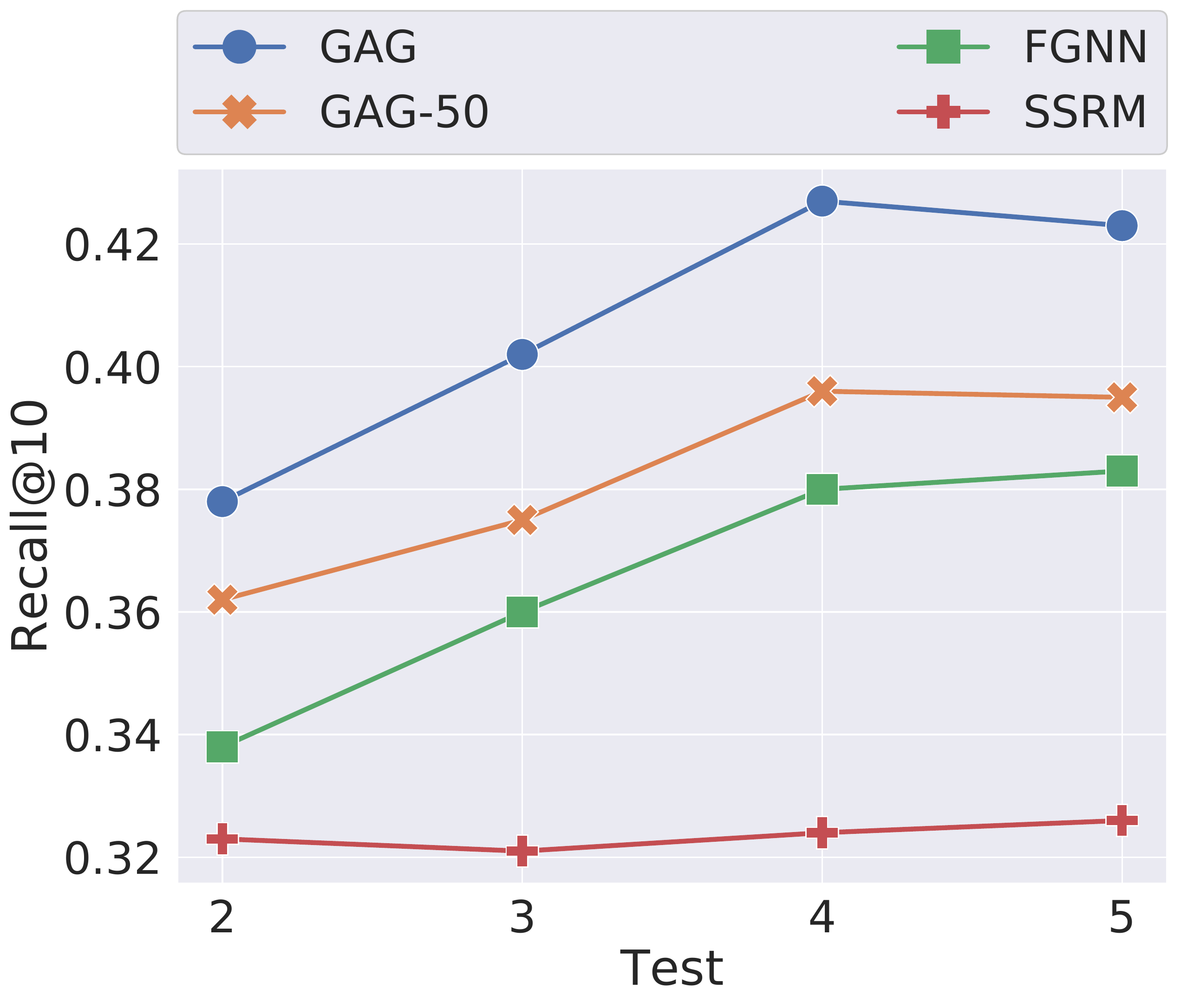}
    }
    \subfigure[MRR@10 index.]{
    \label{fig:mrr10-gowalla}
    \includegraphics[width=0.47\linewidth]{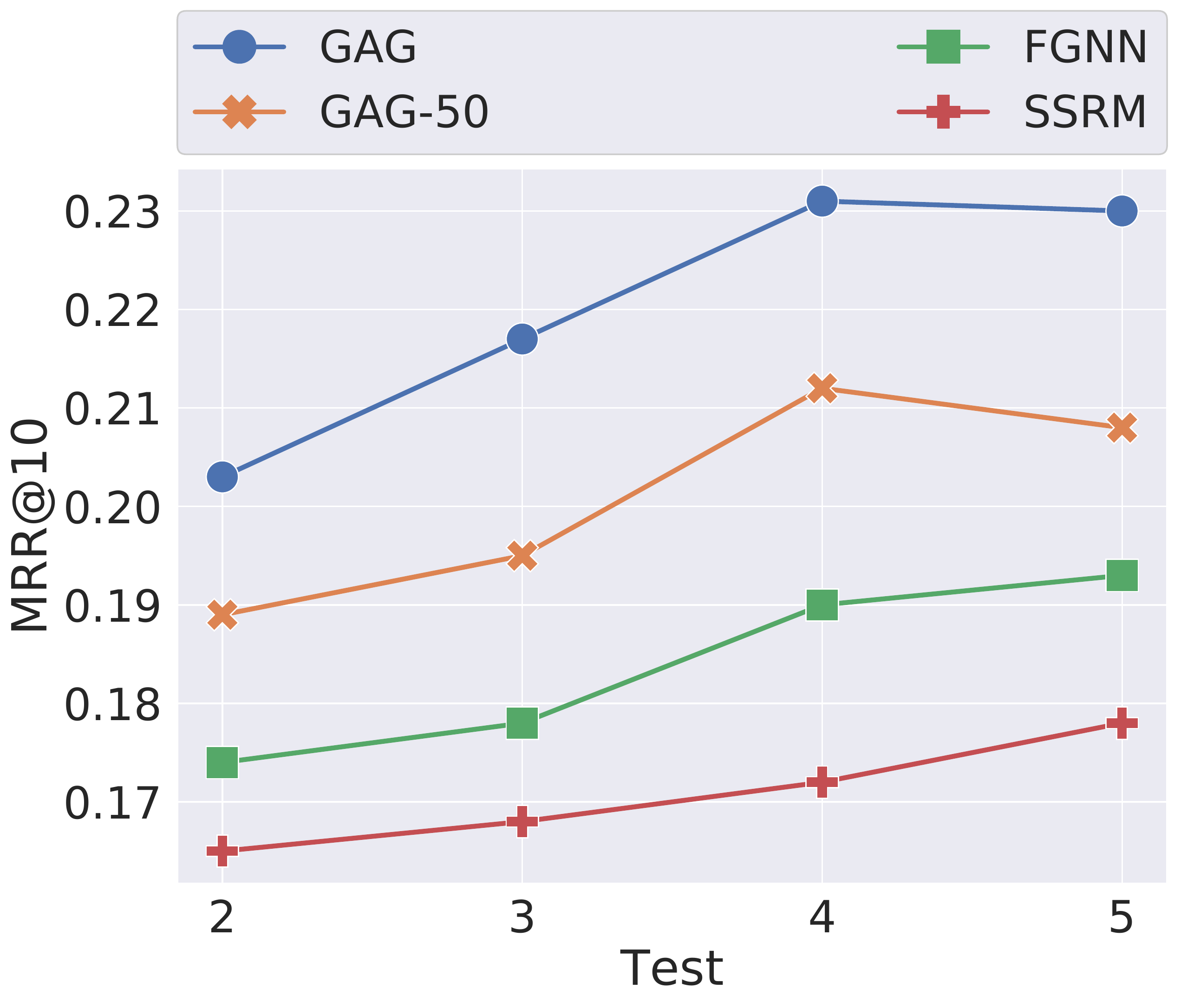}
    }
    \vspace{-0.5cm}
    \caption{In-depth results of streaming session-based recommendation performance on Gowalla.}
    \label{fig:gowalla_indepth}
\end{figure}

Recently, graph neural networks have been demonstrated to have a strong ability to model structured data. For example, FGNN is a state-of-the-art method for SR. In this experiment, we can see that FGNN achieves a comparable performance with SSRM. Compared with GAG-50, FGNN has a worse performance because it is only designed for SR and cannot model the user information.

\subsubsection{\textbf{In-depth Comparison}}
We further evaluate GAG model by the top-5 and top-10 recommendation results on the Gowalla dataset. Specifically, we use the Recall@$K$ and the MRR@$K$ ($K=5,10$) scores to demonstrate the result in Fig.~\ref{fig:gowalla_indepth}.

According to the results, GAG and GAG-50 still have superiority in the higher standard recommendation. Compared with GNN-based methods, SSRM has a greater drop in both top-5 and top-10 performance, implying that the graph structure and GNN are more suitable for the session representation and the generalization ability. In contrast, the attention mechanism fails to distinguish the item transition pattern in sessions.

\subsection{Effect of Global Attribute}
\label{sec:rq2}
In our GAG model, we utilize the global attribute in both the node embedding update and the global attribute update itself. In this experiment, we conduct the ablation study and make different substitutions of the global attribute to evaluate its efficacy. We use the Recall@20 and MRR@20 on Gowalla and LastFM datasets to evaluate the performance.

\begin{figure}[t]
    \centering
    \subfigure[Recall@20 on Gowalla.]{
    \label{fig:r20-gowalla_ga_ablation}
    \includegraphics[width=0.46\linewidth]{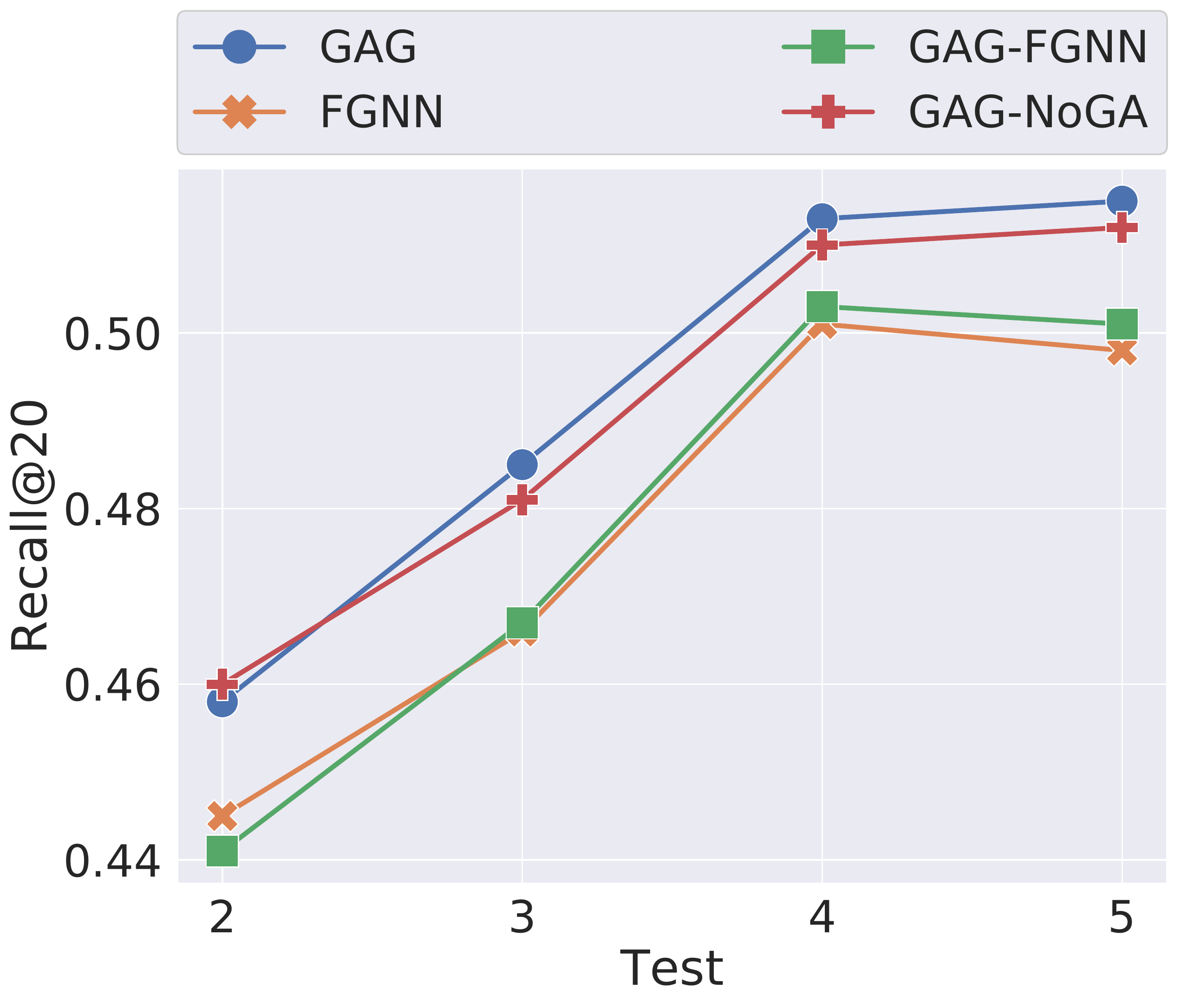}
    }
    \subfigure[MRR@20 on Gowalla.]{
    \label{fig:mrr20-gowalla_ga_ablation}
    \includegraphics[width=0.47\linewidth]{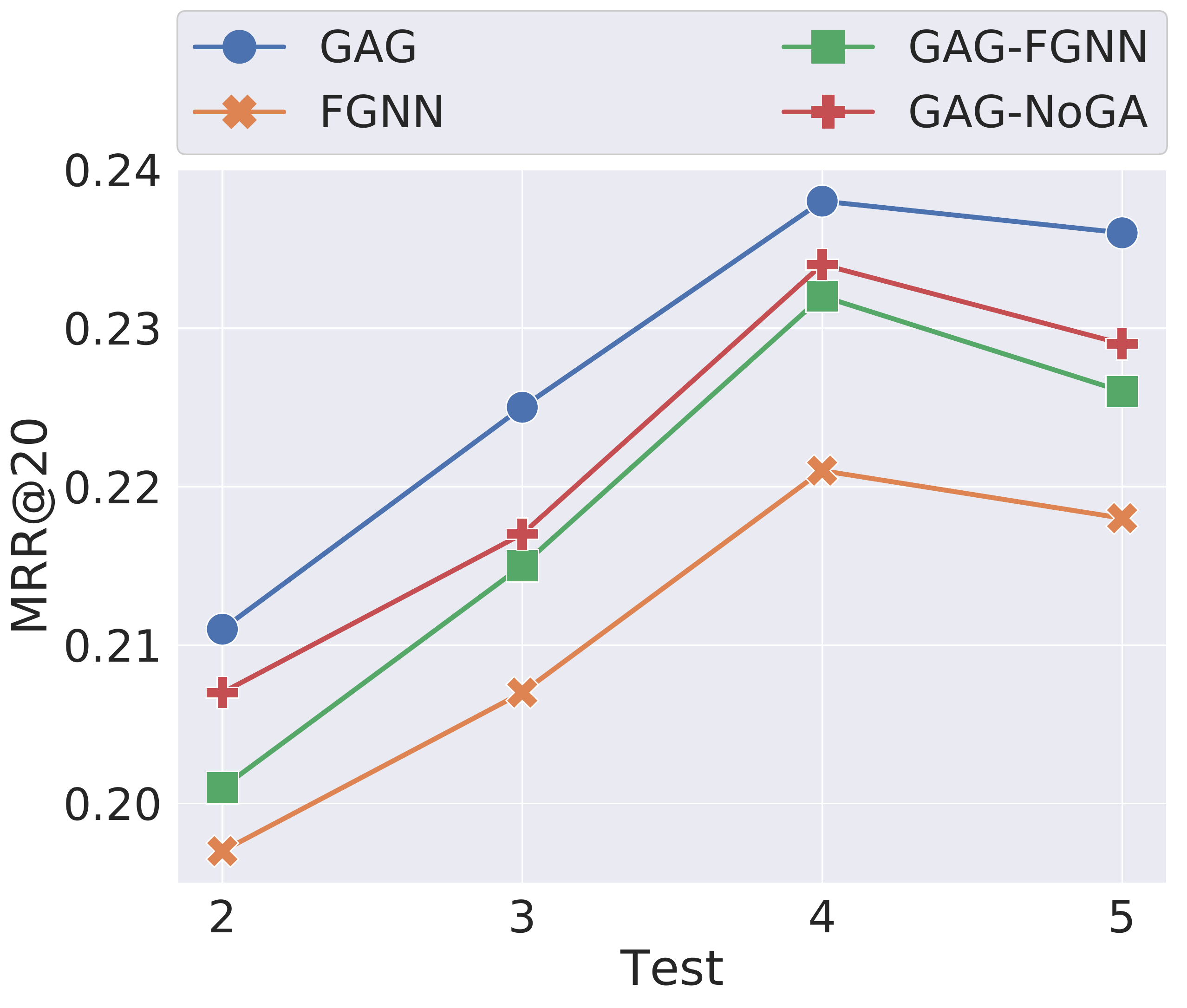}
    }
    \subfigure[Recall@20 on LastFM.]{
    \label{fig:r20-lastfm_ga_ablation}
    \includegraphics[width=0.46\linewidth]{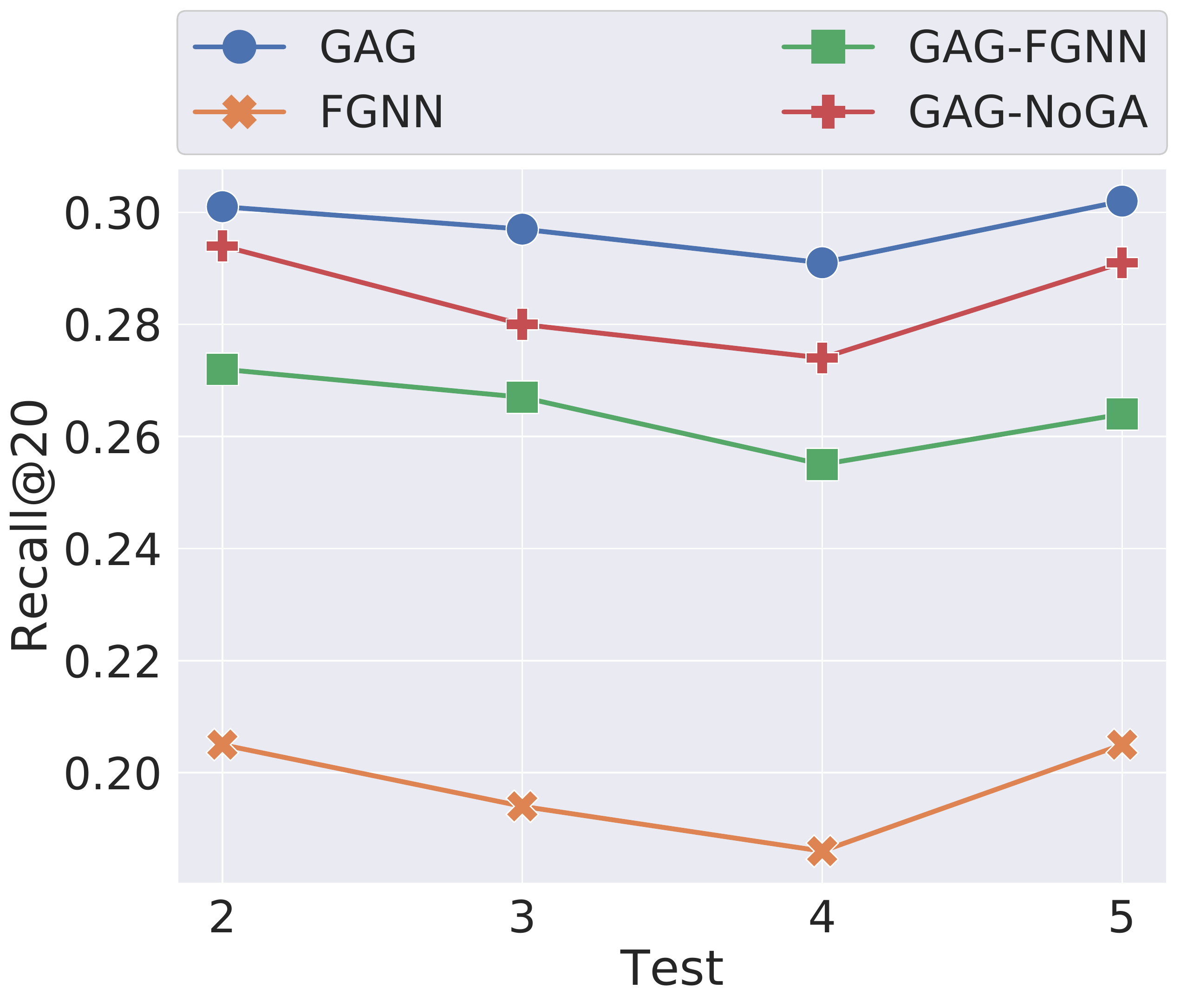}
    }
    \subfigure[MRR@20 on LastFM.]{
    \label{fig:mrr20-lastfm_ga_ablation}
    \includegraphics[width=0.47\linewidth]{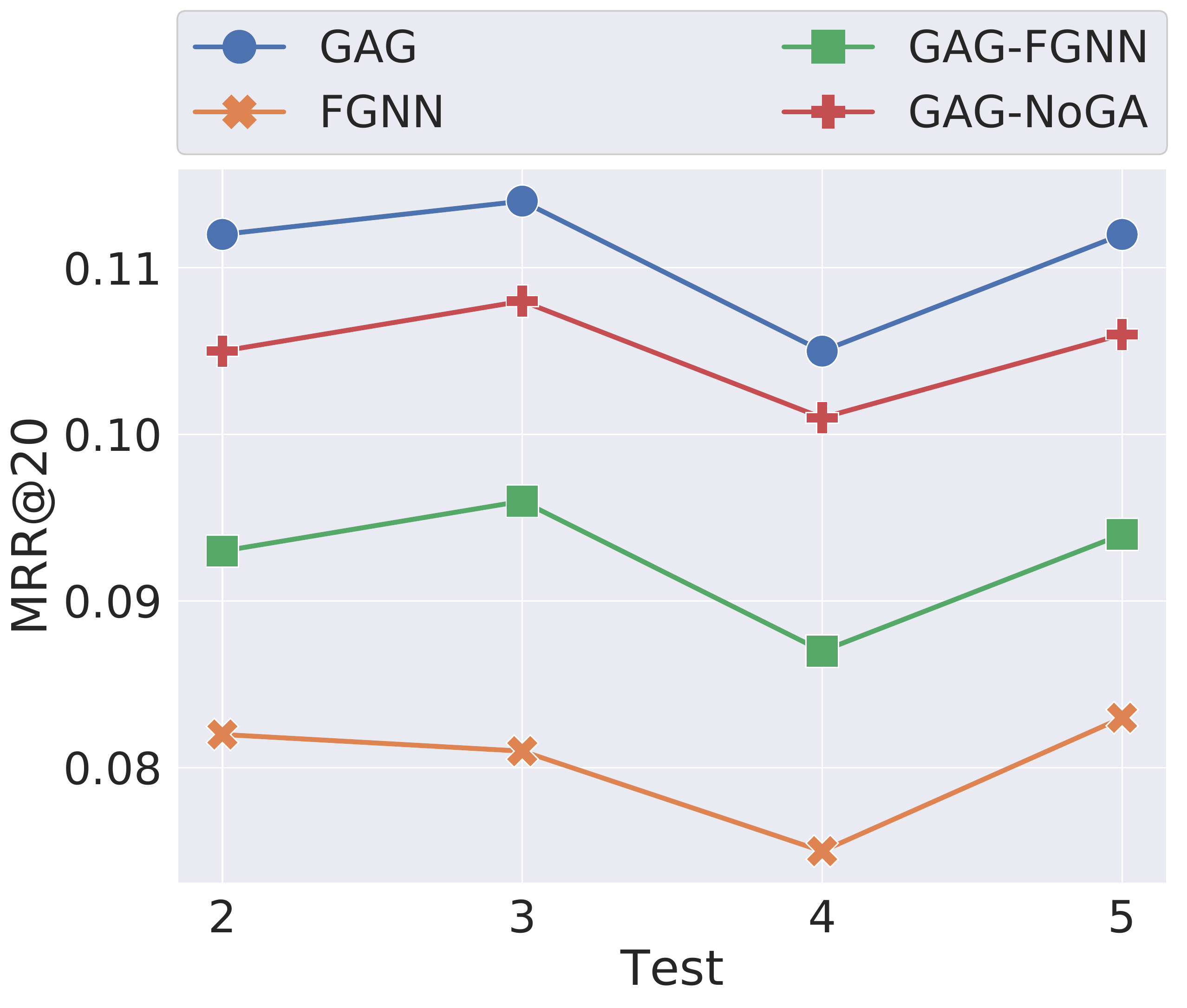}
    }
    \vspace{-0.5cm}
    \caption{Results of the ablation study of the global attribute.}
    \label{fig:ga}
\end{figure}

\subsubsection{\textbf{Ablation Study}}
In this experiment, we compare the GAG model with the following variants:
\begin{itemize}
   \item \textbf{FGNN}: FGNN uses the GNN layer that does not take the user information in both the node update layer and the readout function (readout function in a normal GNN model represents the graph level output function). It serves as the basic baseline method.
   \item \textbf{GAG-FGNN}: We substitute the node update function with FGNN's node update layer and maintain the global attribute update function in GAG to evaluate the integration of the user's information in the global attribute update.
   \item \textbf{GAG-NoGA}: In this variant, we keep the global attribute in the node update procedure while removing it in the global attribute update function.
\end{itemize}

The results are presented in Fig.~\ref{fig:ga}. Each module using the global attribute has a contribution to the recommendation performance. In general, FGNN is the worst because it neglects the global attribute. GAG-FGNN and GAG-NoGA both make improvements by introducing the global attribute. Specifically, GAG-FGNN uses the global attribute in the global attribute update function while GAG-NoGA incorporates the global attribute in the node update function. Comparing these two variants, the results prove that the global attribute applied to the node update procedure has a greater impact on the recommendation performance than in its self-update.

\begin{figure}[t]
    \centering
    \subfigure[Recall@20 on Gowalla.]{
    \label{fig:gowalla_ent_R20}
    \includegraphics[width=0.46\linewidth]{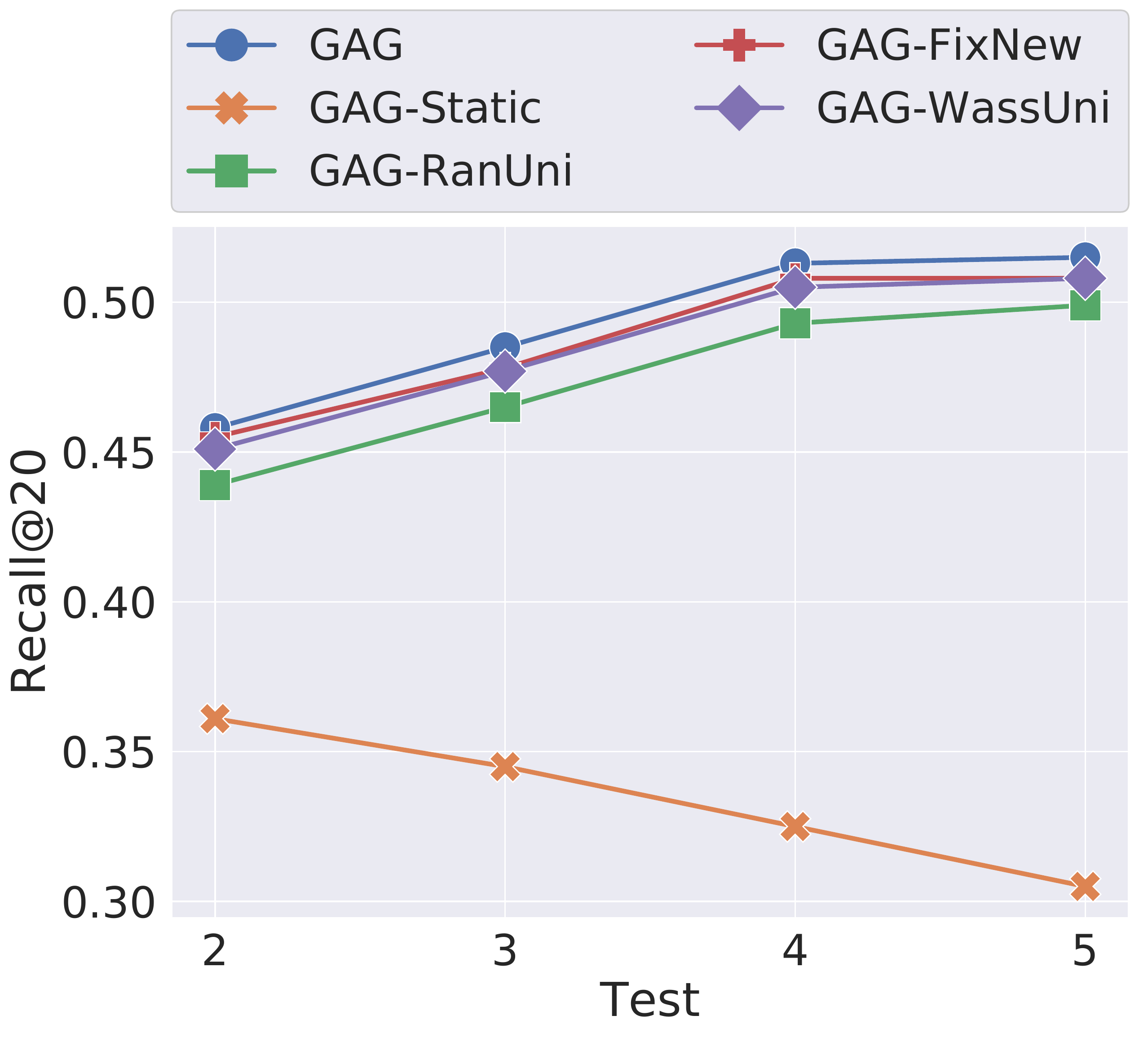}
    }
    \subfigure[MRR@20 on Gowalla.]{
    \label{fig:gowalla_ent_M20}
    \includegraphics[width=0.47\linewidth]{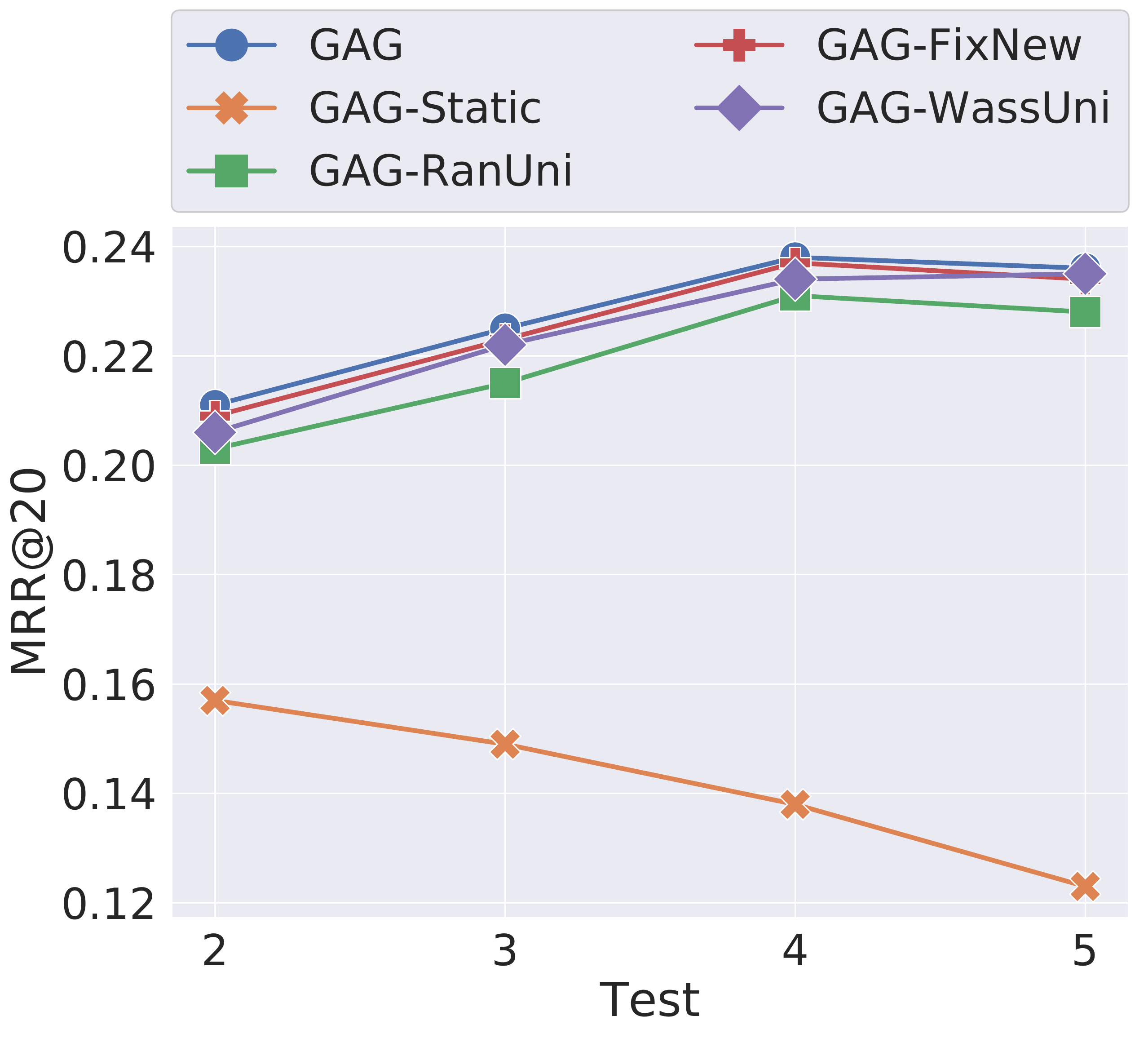}
    }
    \subfigure[Recall@20 on LastFM.]{
    \label{fig:lastfm_ent_R20}
    \includegraphics[width=0.46\linewidth]{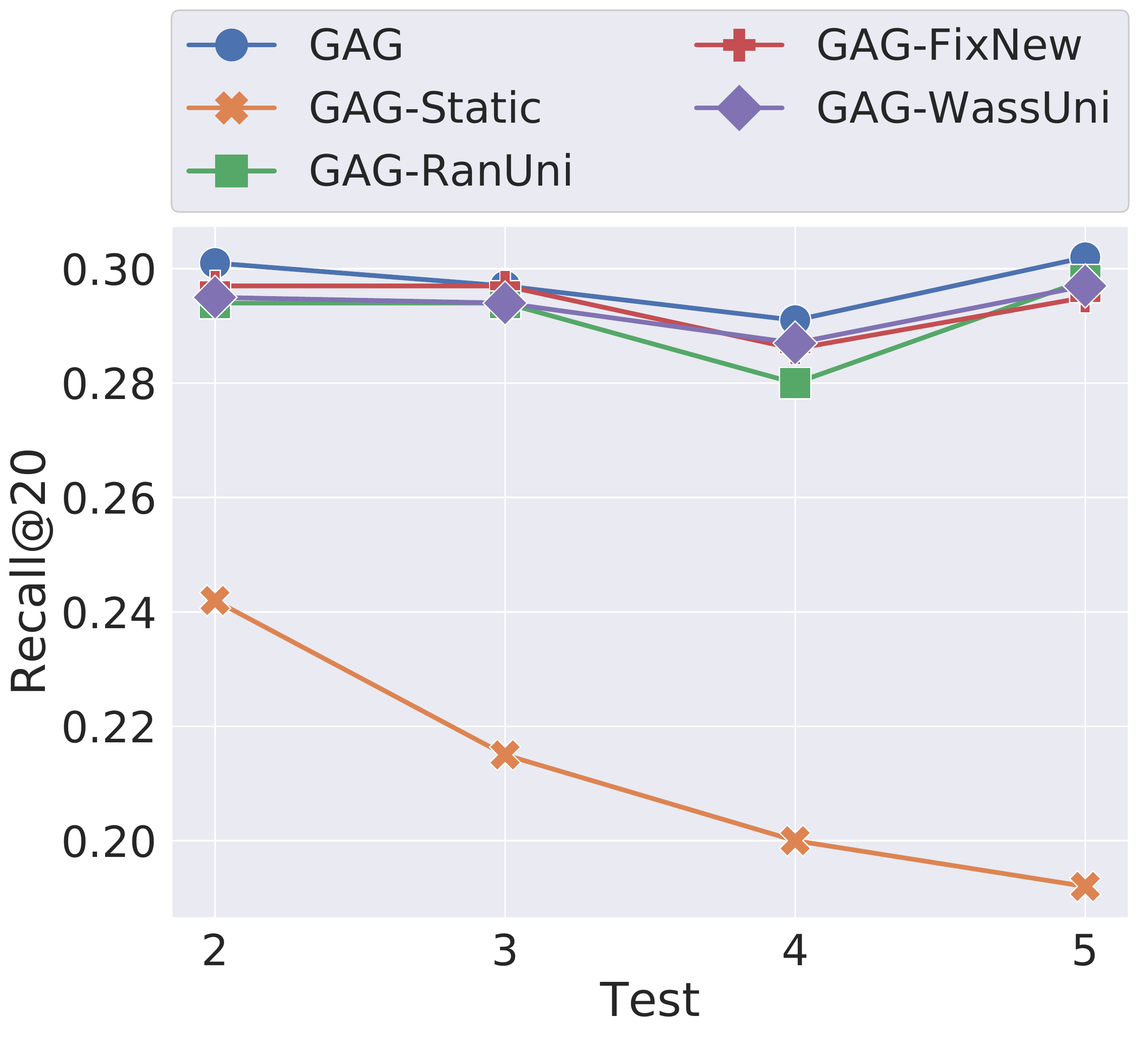}
    }
    \subfigure[MRR@20 on LastFM.]{
    \label{fig:lastfm_ent_M20}
    \includegraphics[width=0.47\linewidth]{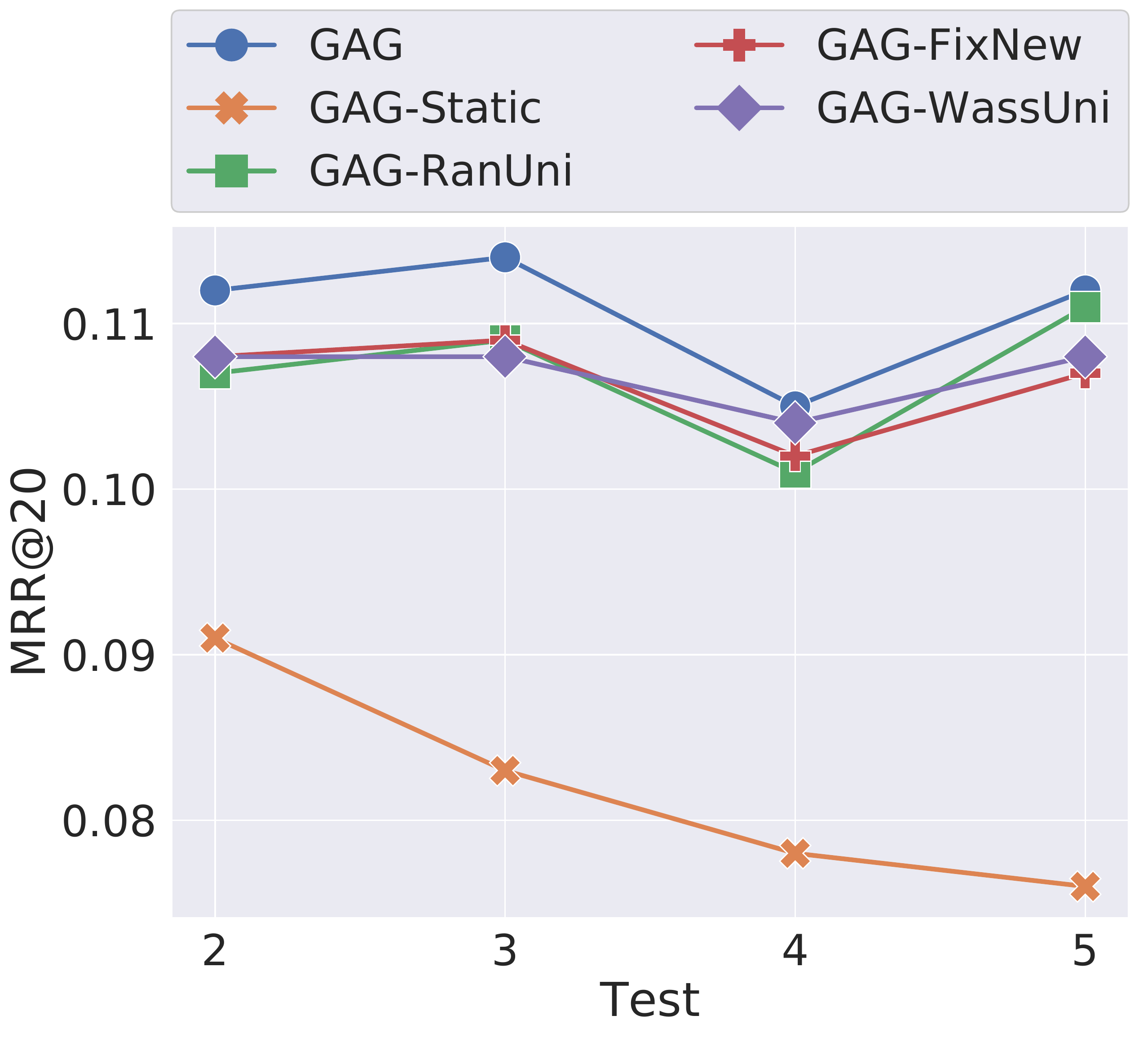}
    }
    \vspace{-0.5cm}
    \caption{Results of the different reservoir sampling strategies.}
    \label{fig:ent}
\end{figure}

\subsection{Effect of Wasserstein Reservoir}
\label{sec:rq3}
In this section, we conduct experiments to prove the efficacy of the Wasserstein reservoir. The reservoir consists of two major designs: (1) The sampling procedure is based on the Wasserstein distance between the session's predictive distribution and the real interaction; (2) Sessions containing new items or new users will be added to the training sample directly. We substitute the Wasserstein reservoir with other reservoirs to evaluate how the design of the reservoir affects the performance of the GAG model in the SSR problem.

\subsubsection{\textbf{Ablation Study}}
In this experiment, an ablation study is conducted to prove the efficacy of both designs in our Wasserstein reservoir. The variants are listed out as follows:
\begin{itemize}
   \item \textbf{GAG-Static}: For this method, we simply eliminate the online training of the model.
   \item \textbf{GAG-RanUni}: This variant only performs random sampling on the union set of the current reservoir and new arrival sessions. It is the most common design of a reservoir.
   \item \textbf{GAG-FixNew}: This variant directly adds a new session containing new items or new users to the training data. For the rest, it still performs a random sampling.
   \item \textbf{GAG-WassUni}: This method samples the training data from the union set of the current reservoir and the new sessions according to their Wasserstein distance.
\end{itemize}

According to the result in Fig.~\ref{fig:ent}, our proposed Wasserstein reservoir achieves the best performance in all situations. For the static recommendation version, GAG-Static, its performance decreases along with the time because there is a shift of the users' preference and the streaming sessions contain new items and new users. In most cases, the random sampling version variant, GAG-RanUni, performs worse than other methods that use a specialized reservoir sampling strategy. The conclusion can be drawn from these two figures that incorporating the sessions containing new items and new users helps with the online update of the model. Comparing GAG-FixNew with GAG-RanUni, the performance of GAG-FixNew is better on Gowalla while it has a decrease in the long-term prediction of $\mathcal{D}^{test,4}$ and $\mathcal{D}^{test,5}$ on LastFM. The model is distracted because the new items and new users in these two parts of the dataset are not representative. Comparing the complete GAG model with GAG-WassUni, it can be seen that the incorporation of new users and items can help with online training. Furthermore, the efficacy of Wasserstein distance is demonstrated. With the Wasserstein distance-based sampling, GAG-WassUni outperforms GAG-RanUni in most cases. Similarly, GAG also has higher scores than GAG-FixNew, which randomly samples the training data.

\begin{figure}[t]
    \centering
    \subfigure[Recall@20 on Gowalla.]{
    \label{fig:gowalla_res_R20}
    \includegraphics[width=0.46\linewidth]{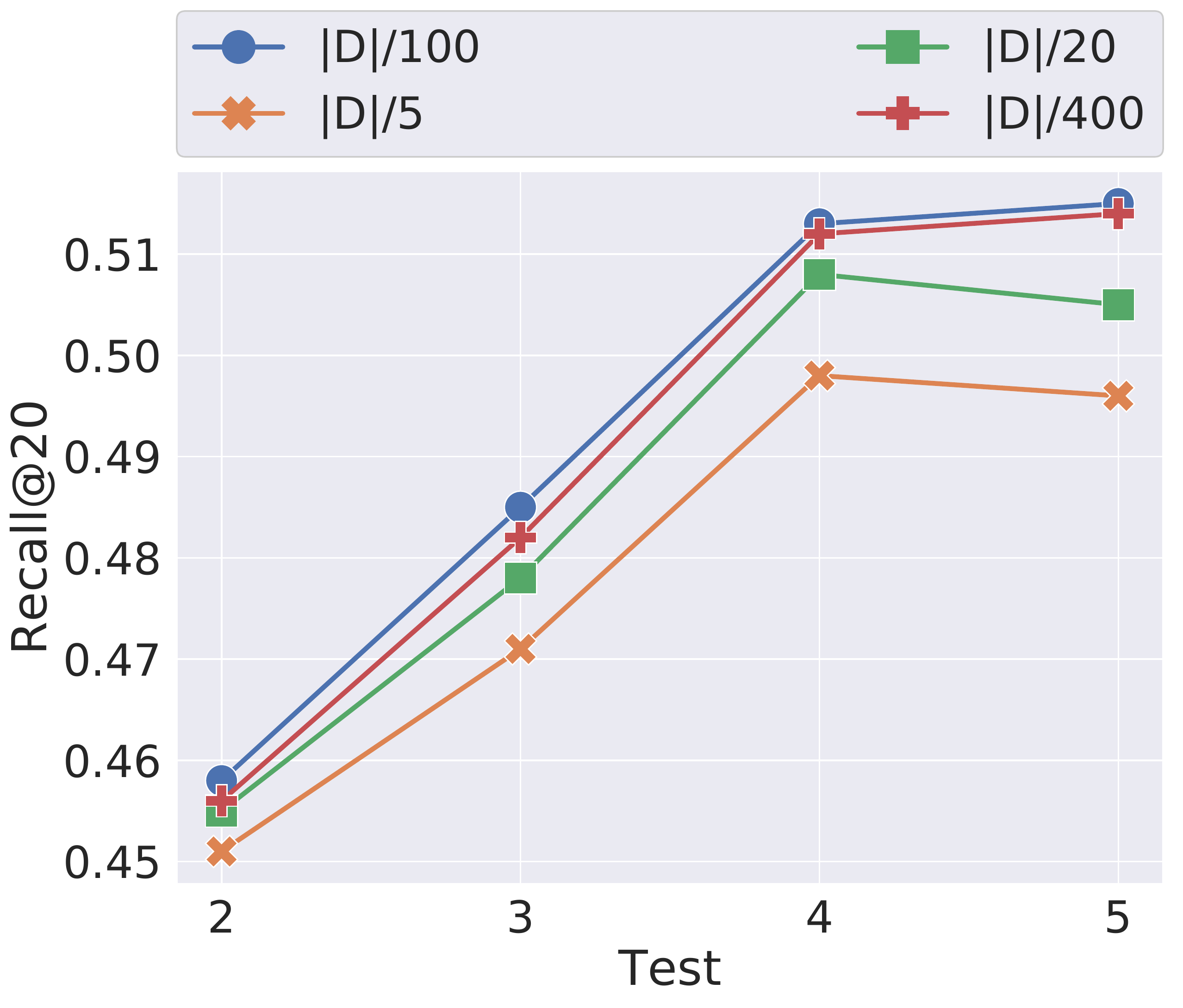}
    }
    \subfigure[MRR@20 on Gowalla.]{
    \label{fig:gowalla_res_M20}
    \includegraphics[width=0.47\linewidth]{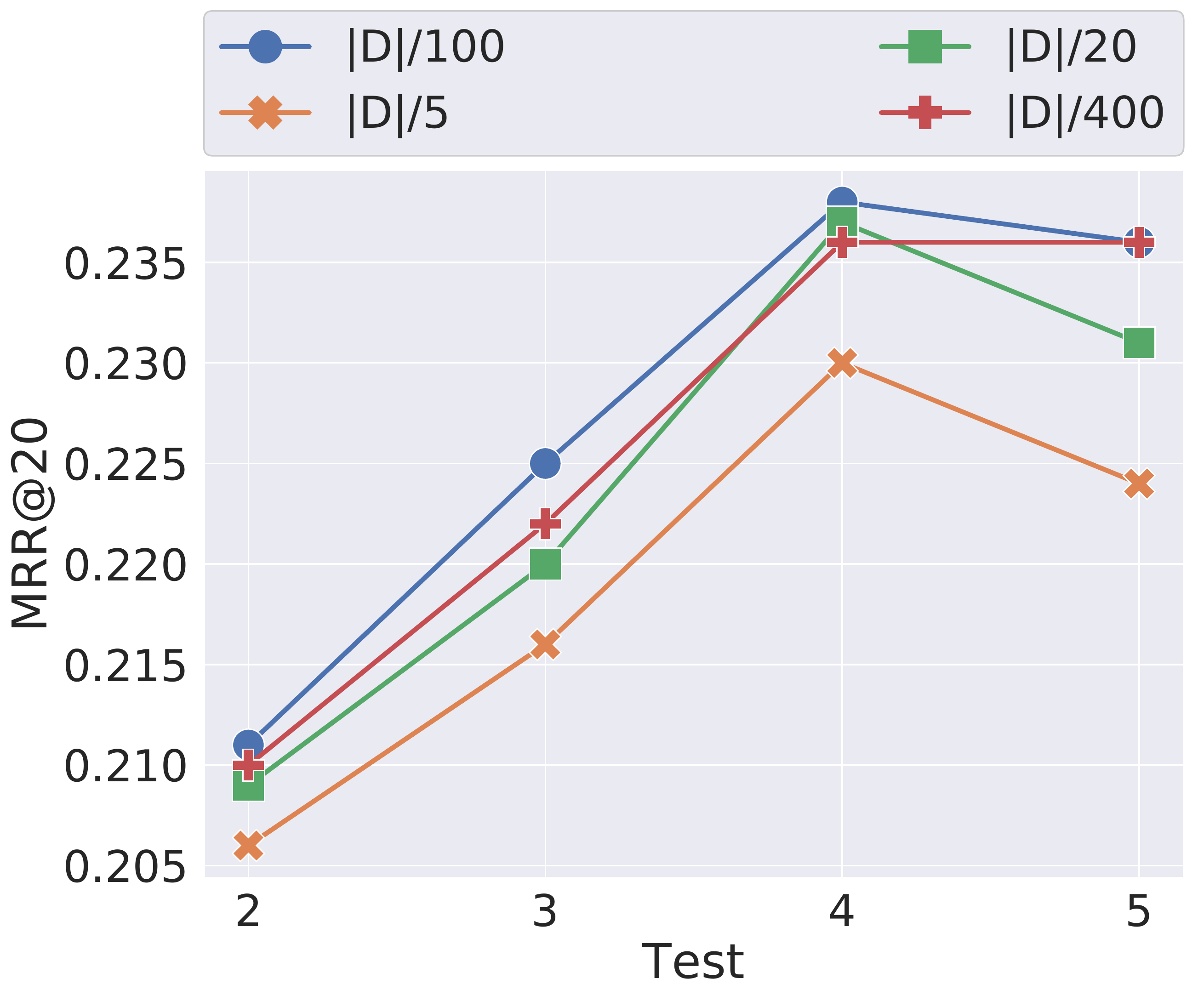}
    }
    \subfigure[Recall@20 on LastFM.]{
    \label{fig:lastfm_res_R20}
    \includegraphics[width=0.46\linewidth]{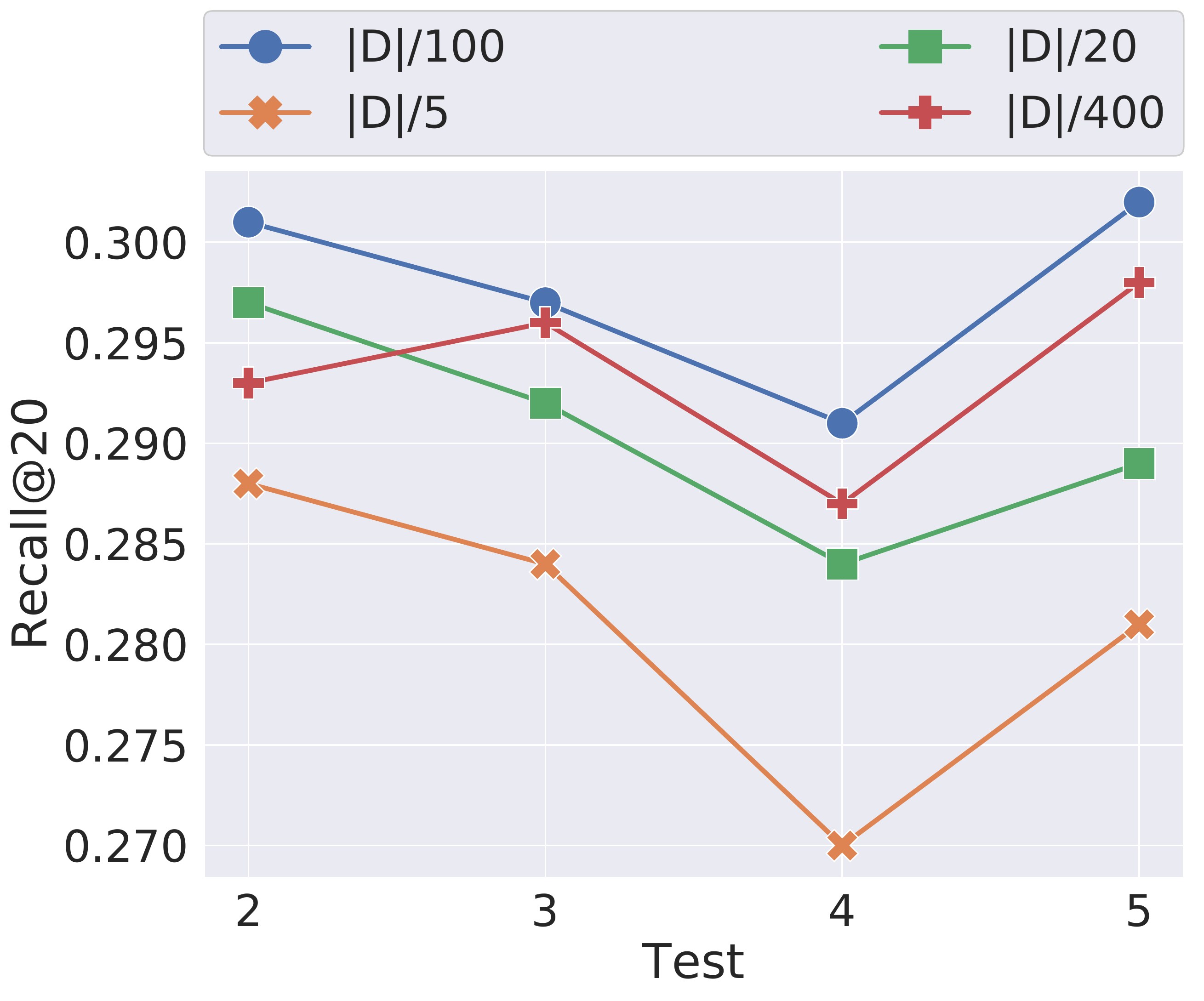}
    }
    \subfigure[MRR@20 on LastFM.]{
    \label{fig:lastfm_res_M20}
    \includegraphics[width=0.47\linewidth]{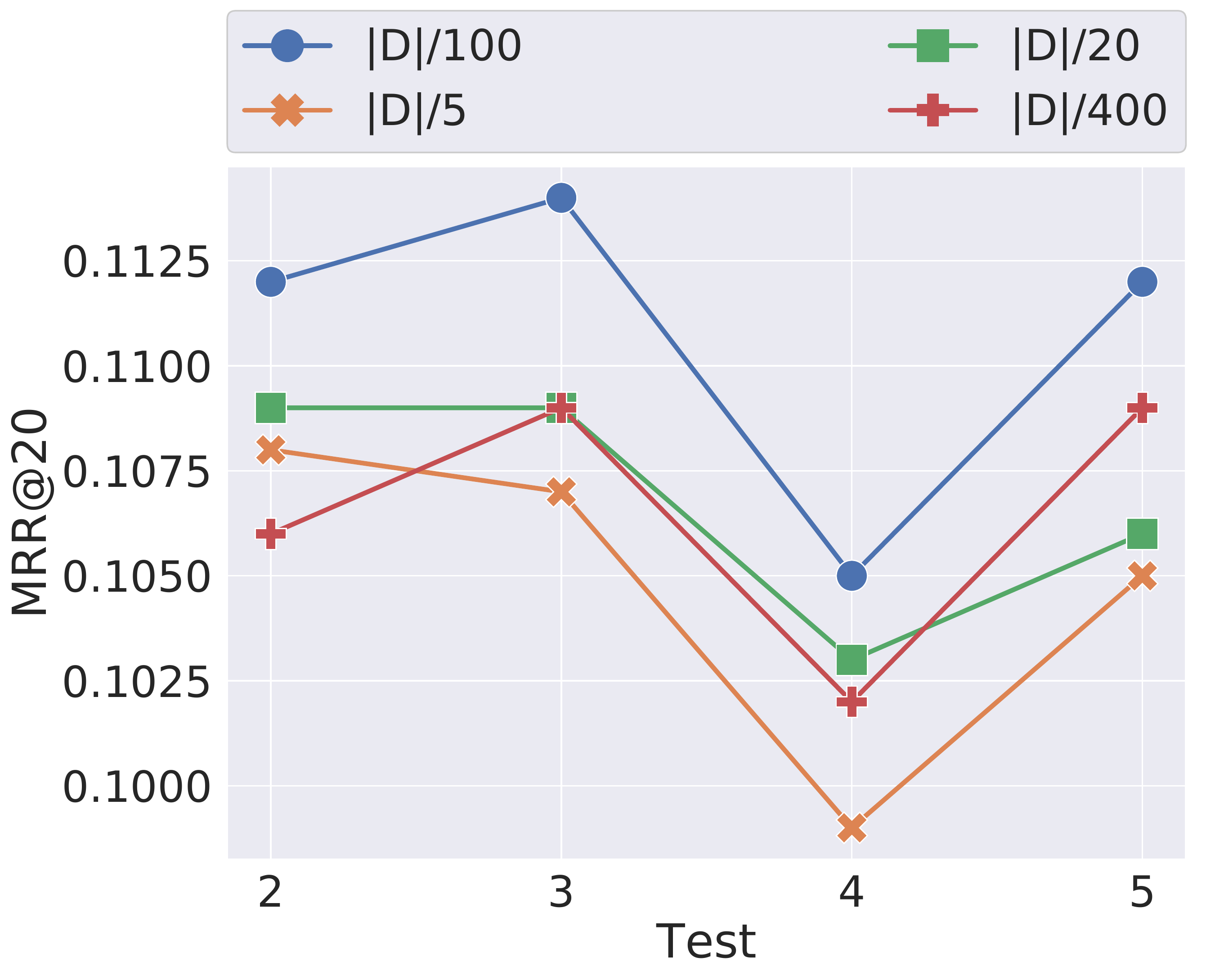}
    }
    \vspace{-0.5cm}
    \caption{Results of different reservoir size.}
    \label{fig:res_size}
\end{figure}

\subsubsection{\textbf{Reservoir Efficiency}}
There are two important parameters in the design of the Wasserstein reservoir: the reservoir size and the window size. On one hand, the reservoir size indicates the volume of the reservoir, which determines the storage requirement of the online update for the recommender system. On the other hand, the window size restricts how many data instances will be sampled for the online training, which represents the work load of the online update for the recommender system.

The default reservoir size is set to $|D|/100$. For comparison, we change the reservoir size to $\{|D|/5,|D|/20,|D|/400\}$ to evaluate the effect of the reservoir size. Results of different reservoir sizes are presented in Fig.~\ref{fig:res_size}. When the reservoir size is set to $|D|/100$, our GAG model achieves the best performance. When the reservoir size increases, the probability of the new sessions stored in the reservoir decreases, which makes the model concentrate more on the historical data. However, for the streaming data, the recent ones are more representative of the users' recent preference. When the reservoir size decreases, the streaming performance drops on a smaller scale, which indicates that new sessions are more important for the recommendation performance. For the state-of-the-art method SSRM, it achieves its highest performance with the reservoir size set to $|D|/20$, which is 5 times larger than our GAG model. Apparently, our design has a higher efficiency in reservoir storage.

\begin{figure}[t]
    \centering
    \subfigure[Recall@20 on Gowalla.]{
    \label{fig:gowalla_embed_R20}
    \includegraphics[width=0.46\linewidth]{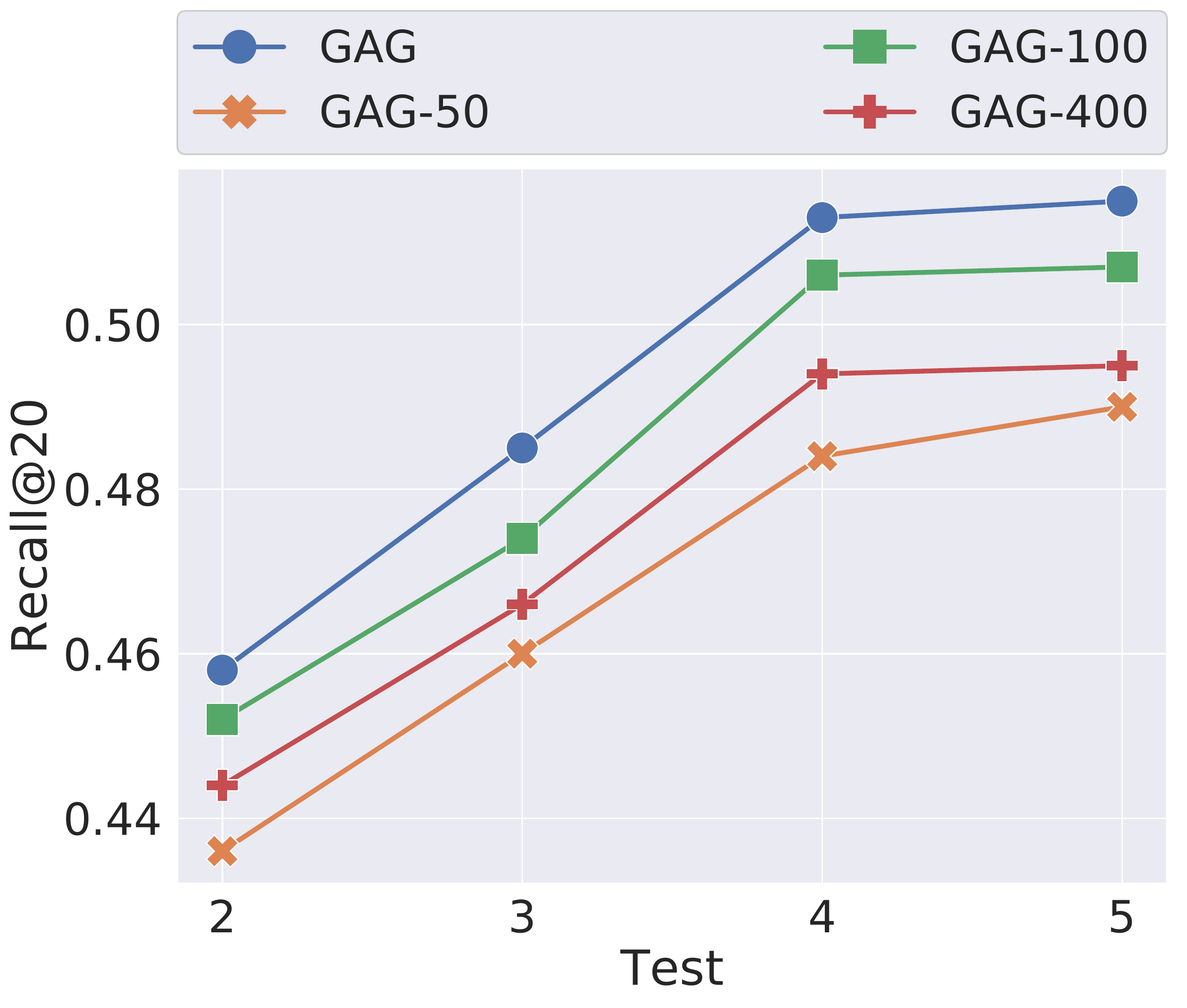}
    }
    \subfigure[MRR@20 on Gowalla.]{
    \label{fig:gowalla_embed_M20}
    \includegraphics[width=0.47\linewidth]{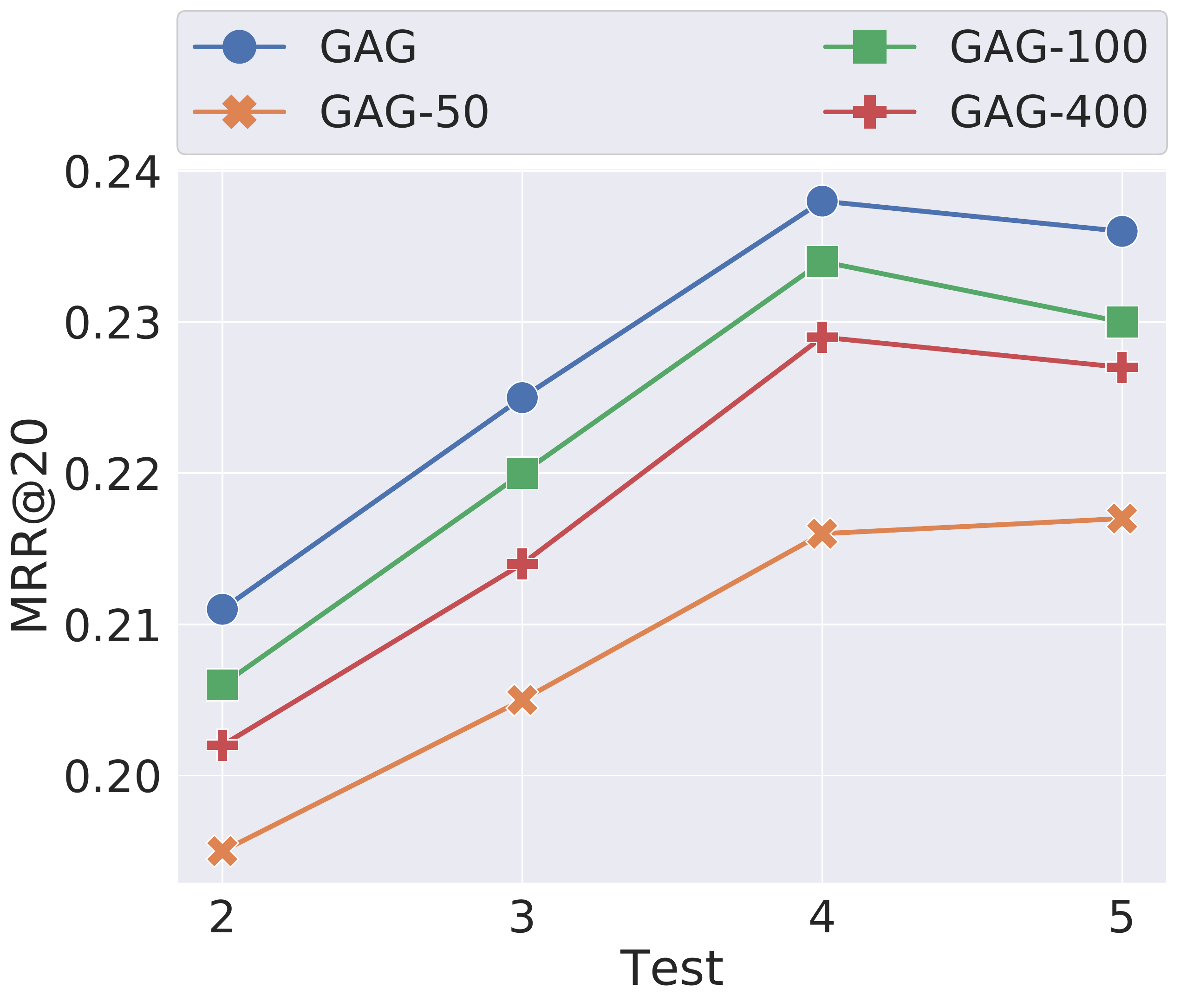}
    }
    \subfigure[Recall@20 on LastFM.]{
    \label{fig:lastfm_embed_R20}
    \includegraphics[width=0.46\linewidth]{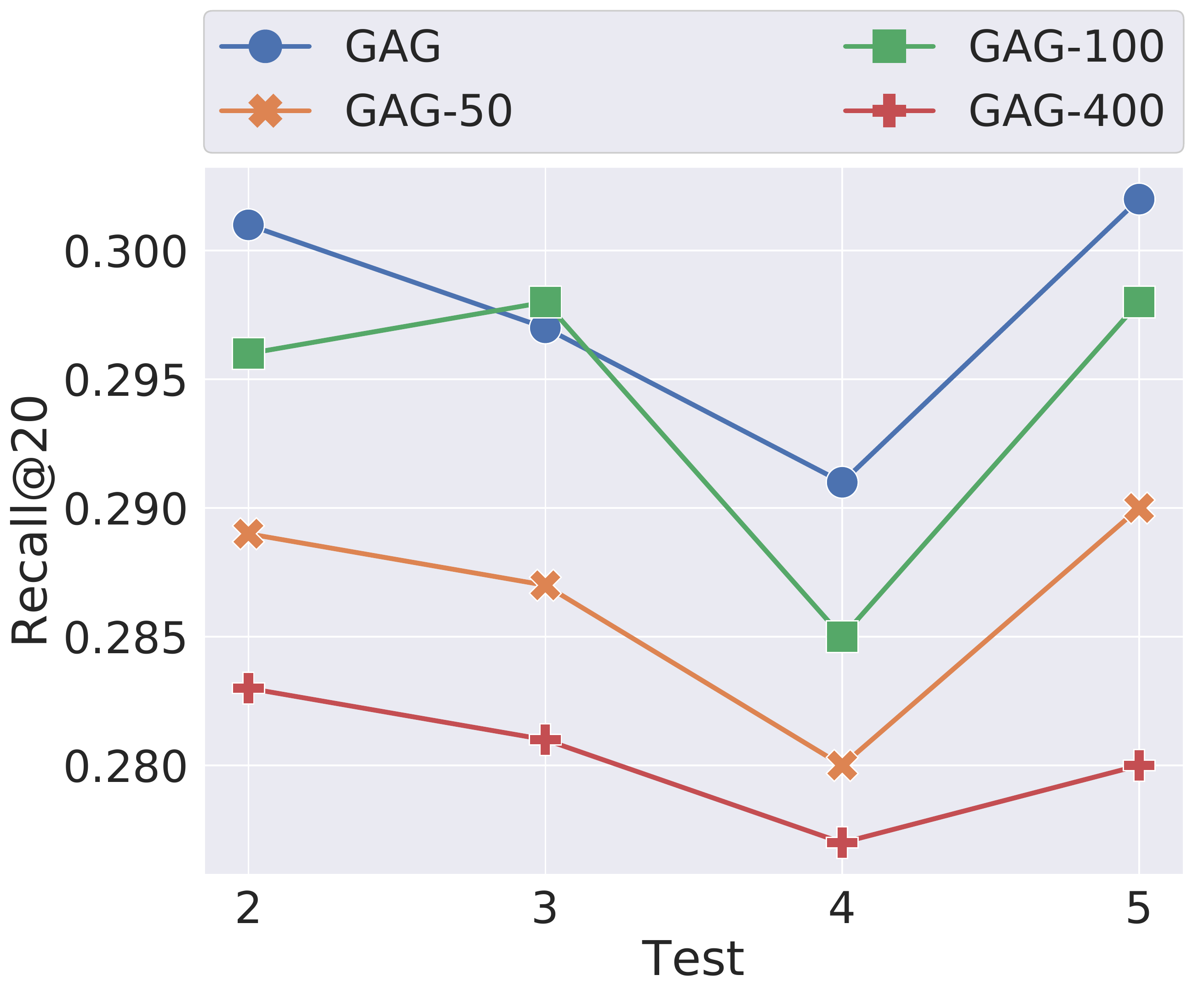}
    }
    \subfigure[MRR@20 on LastFM.]{
    \label{fig:lastfm_embed_M20}
    \includegraphics[width=0.47\linewidth]{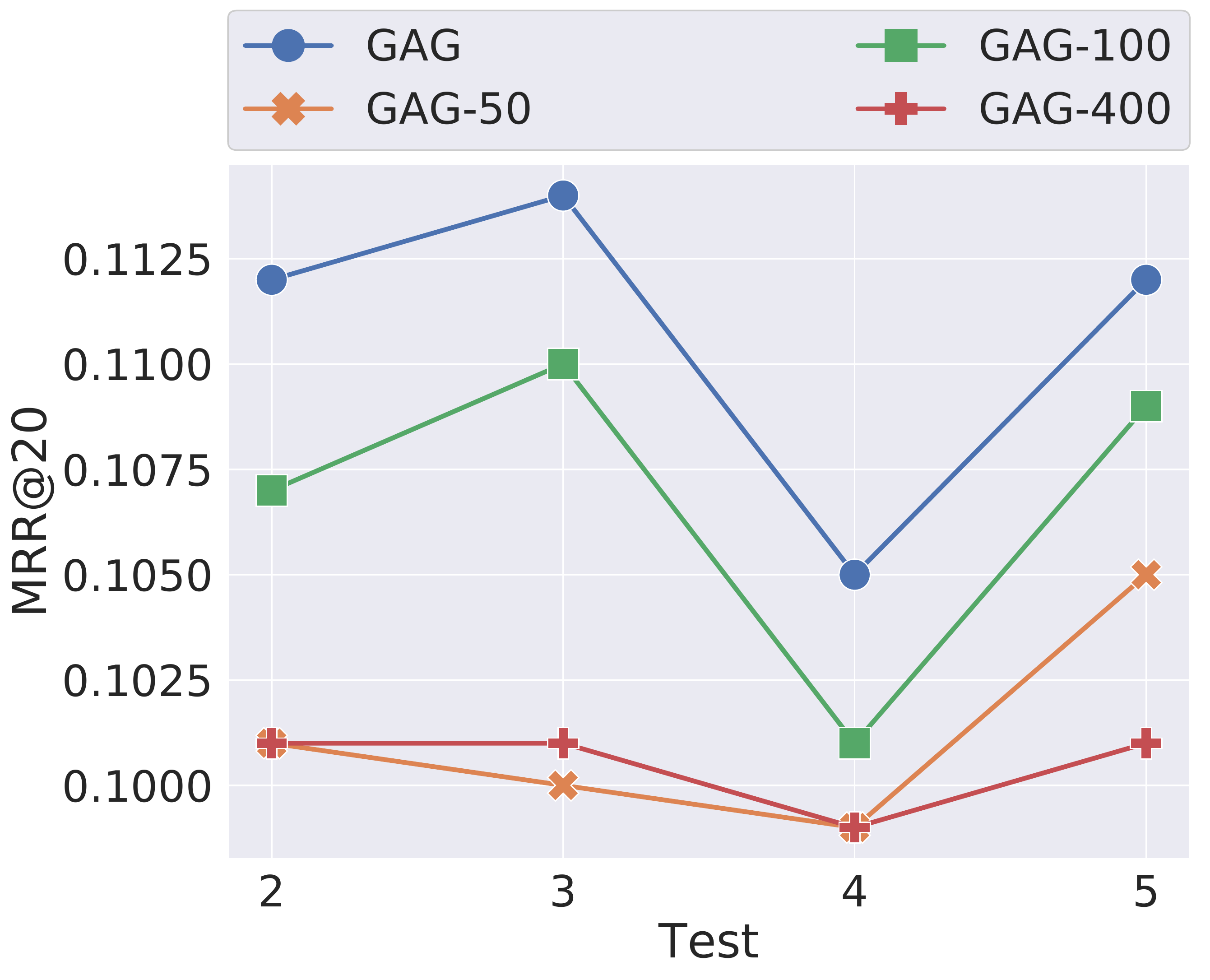}
    }
    \vspace{-0.5cm}
    \caption{Results of different embedding sizes.}
    \label{fig:embed}
\end{figure}

\begin{figure}[t]
    \centering
    \subfigure[Recall@20 on Gowalla.]{
    \label{fig:gowalla_win_R20}
    \includegraphics[width=0.46\linewidth]{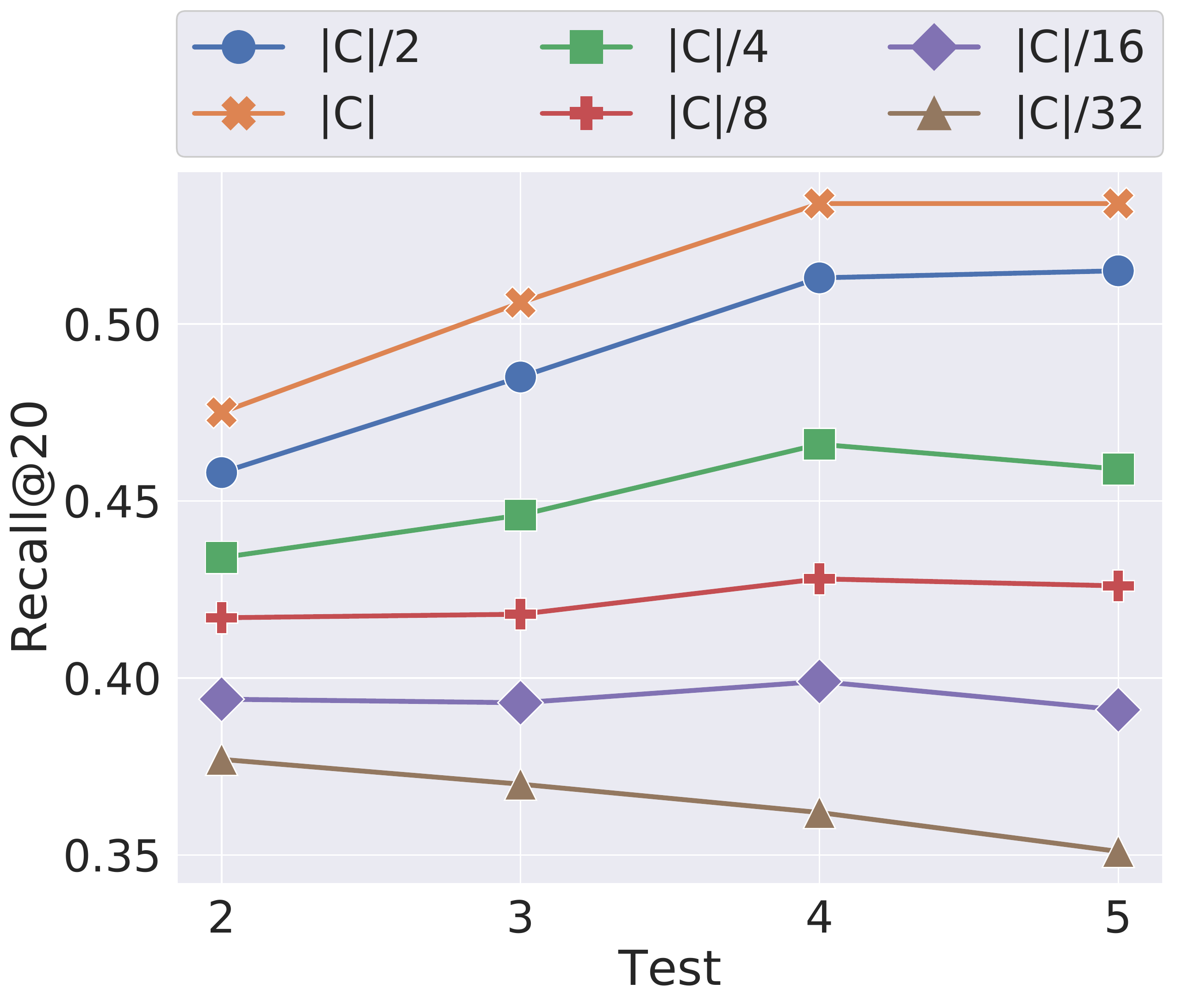}
    }
    \subfigure[MRR@20 on Gowalla.]{
    \label{fig:gowalla_win_M20}
    \includegraphics[width=0.47\linewidth]{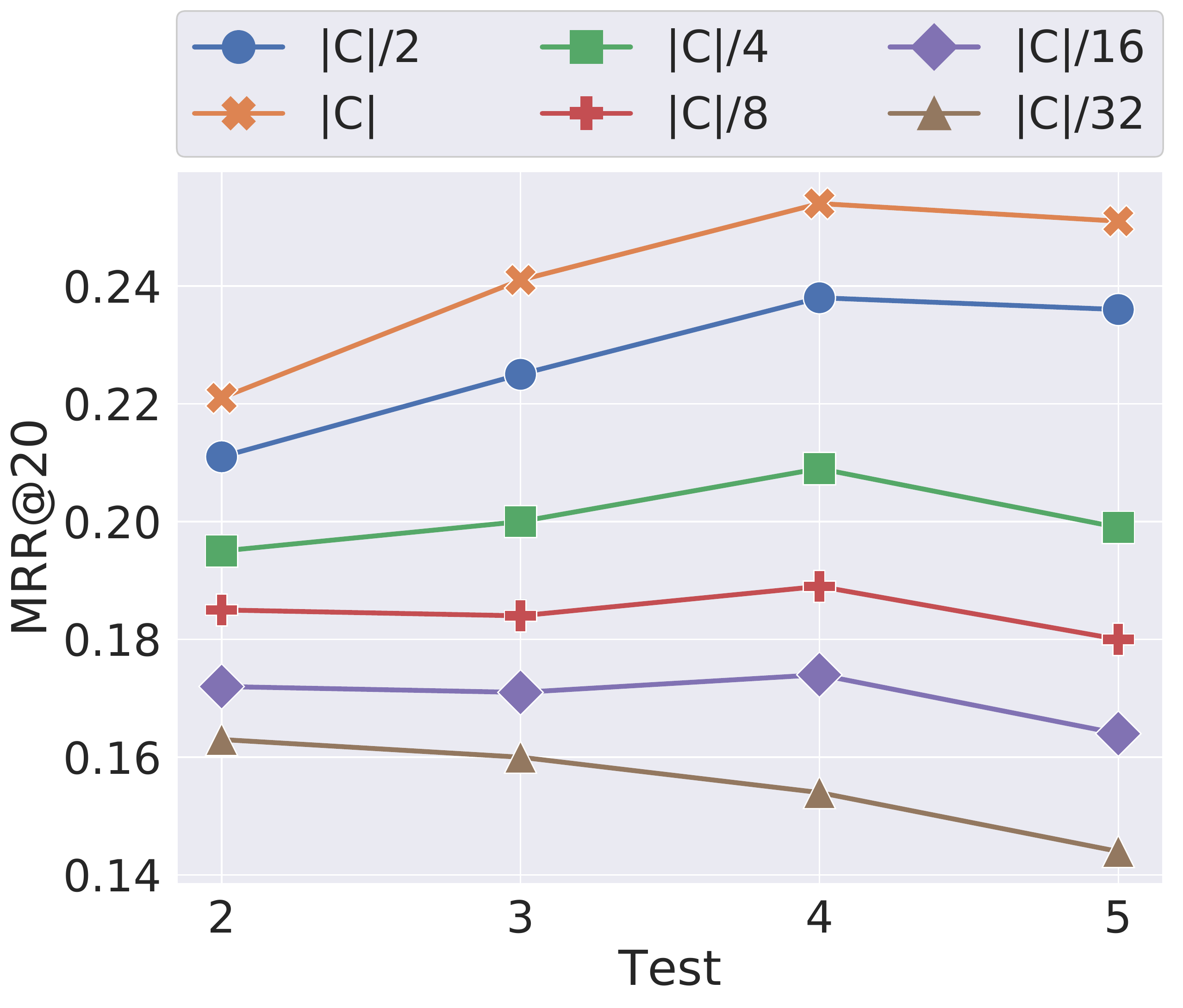}
    }
    \subfigure[Recall@20 on LastFM.]{
    \label{fig:lastfm_win_R20}
    \includegraphics[width=0.46\linewidth]{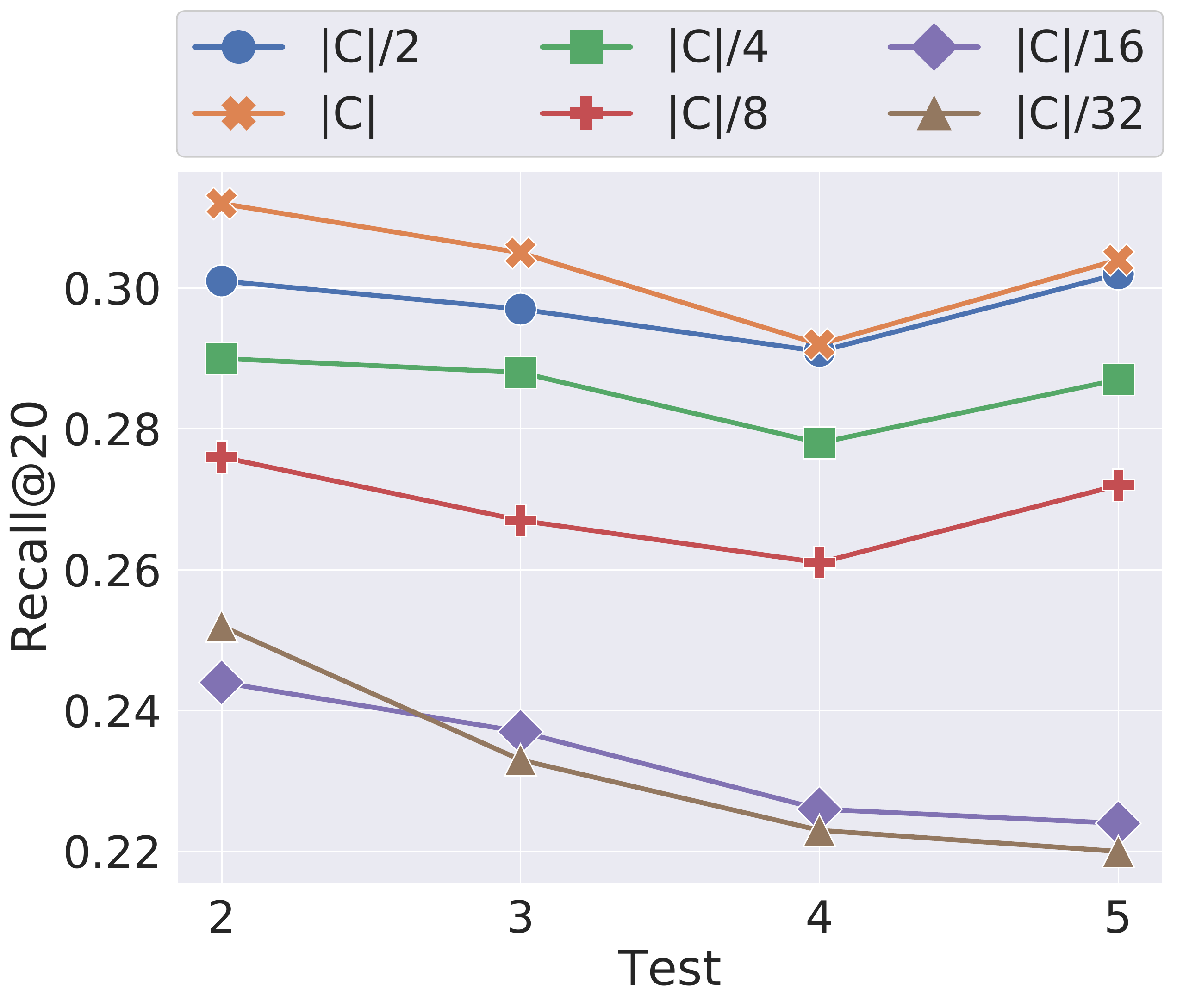}
    }
    \subfigure[MRR@20 on LastFM.]{
    \label{fig:lastfm_win_M20}
    \includegraphics[width=0.47\linewidth]{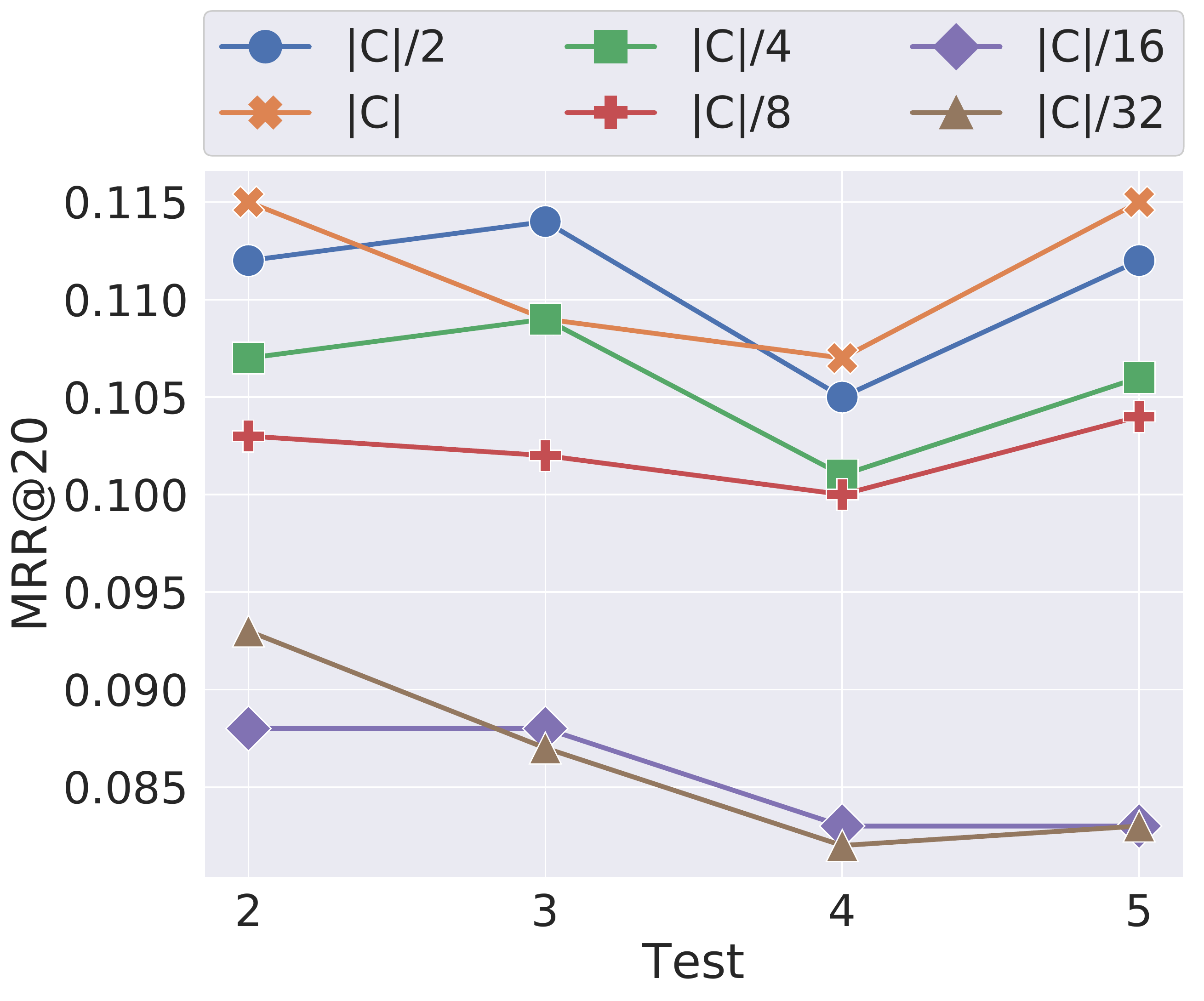}
    }
    \vspace{-0.5cm}
    \caption{Results of different window size.}
    \label{fig:win_size}
\end{figure}

To evaluate the effect of the window size, we substitute the default window size, $|C|/2$, with $\{|C|,|C|/4,|C|/8,|C|/16,|C|/32\}$ to evaluate the effect of the window size. In Fig.~\ref{fig:win_size}, we demonstrate the results. It is clear that when the window size is larger, the model can achieve a better recommendation performance because it can utilize more data to update itself.

\subsection{\textbf{Parameter Sensitivity}}
\label{sec:rq4}
In this section, we conduct experiments to evaluate the parameter sensitivity of our GAG model.

\subsubsection{\textbf{Embedding Size}}
The previous methods achieve the best results when the embedding size is set to 50 or 100. Therefore, we test the following variants of our GAG model with the embedding size of $[50,100,200,400]$: \textbf{GAG-50}, \textbf{GAG-100}, \textbf{GAG} and \textbf{GAG-400}.

In Fig.~\ref{fig:embed}, results of the sensitivity of the embedding size are presented. It is clear that when the embedding size is set to 200, the GAG model has the highest performance in all situations. Size 100 is a relatively strong variant when compared with 50 and 400. When the embedding size is set as 50 and 400, they are unrepresentative and over-parameterized in their respective methods, which causes difficulty in training a strong model.

\subsubsection{Number of Layers}
The number of GAG layers controls the depth of the model. We test our model with different numbers of layers of $[1,2,3]$: \textbf{GAG}, \textbf{GAG-2} and \textbf{GAG-3}.

\begin{figure}[t]
    \centering
    \subfigure[Recall@20 on Gowalla.]{
    \label{fig:gowalla_num_R20}
    \includegraphics[width=0.46\linewidth]{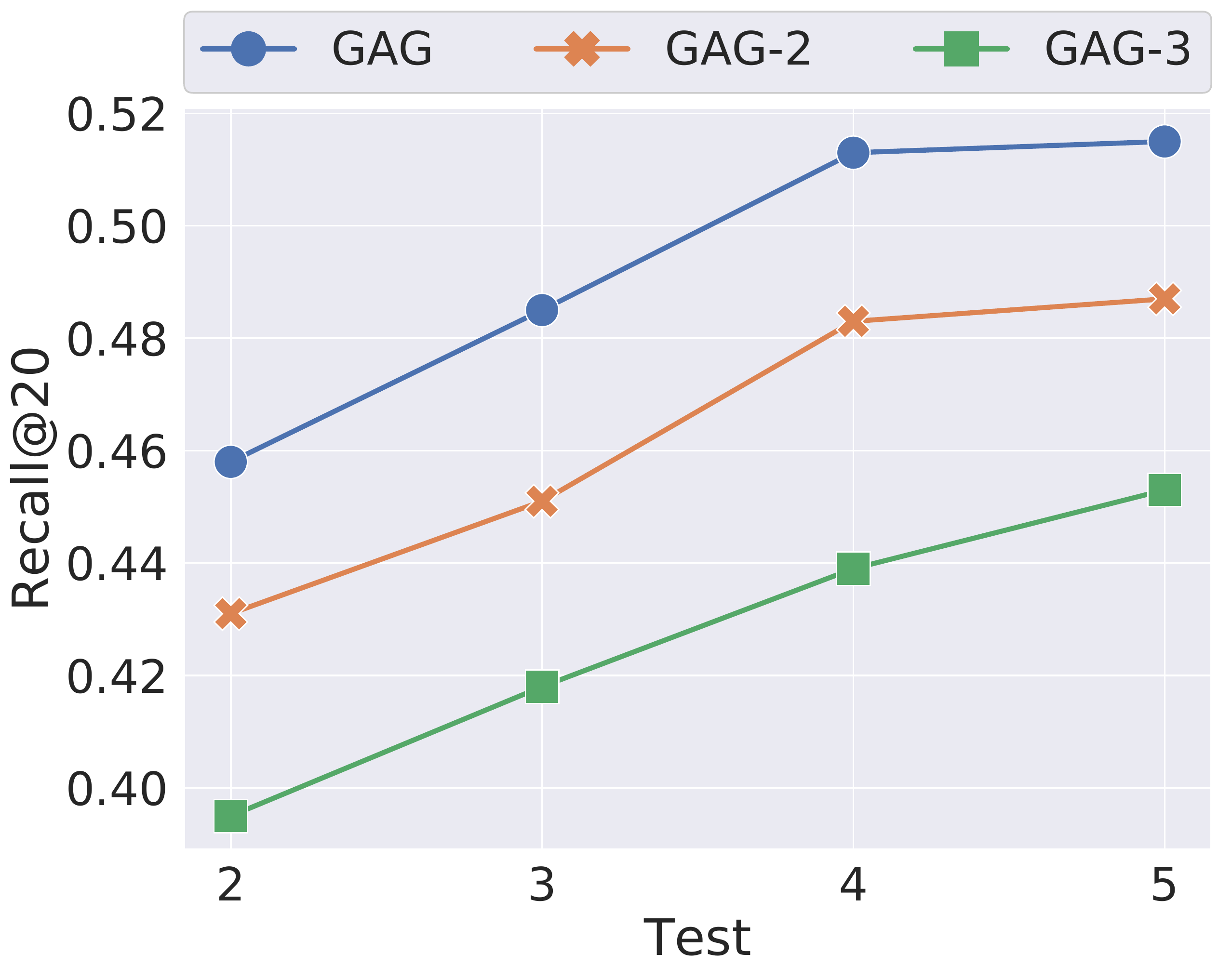}
    }
    \subfigure[MRR@20 on Gowalla.]{
    \label{fig:gowalla_num_M20}
    \includegraphics[width=0.47\linewidth]{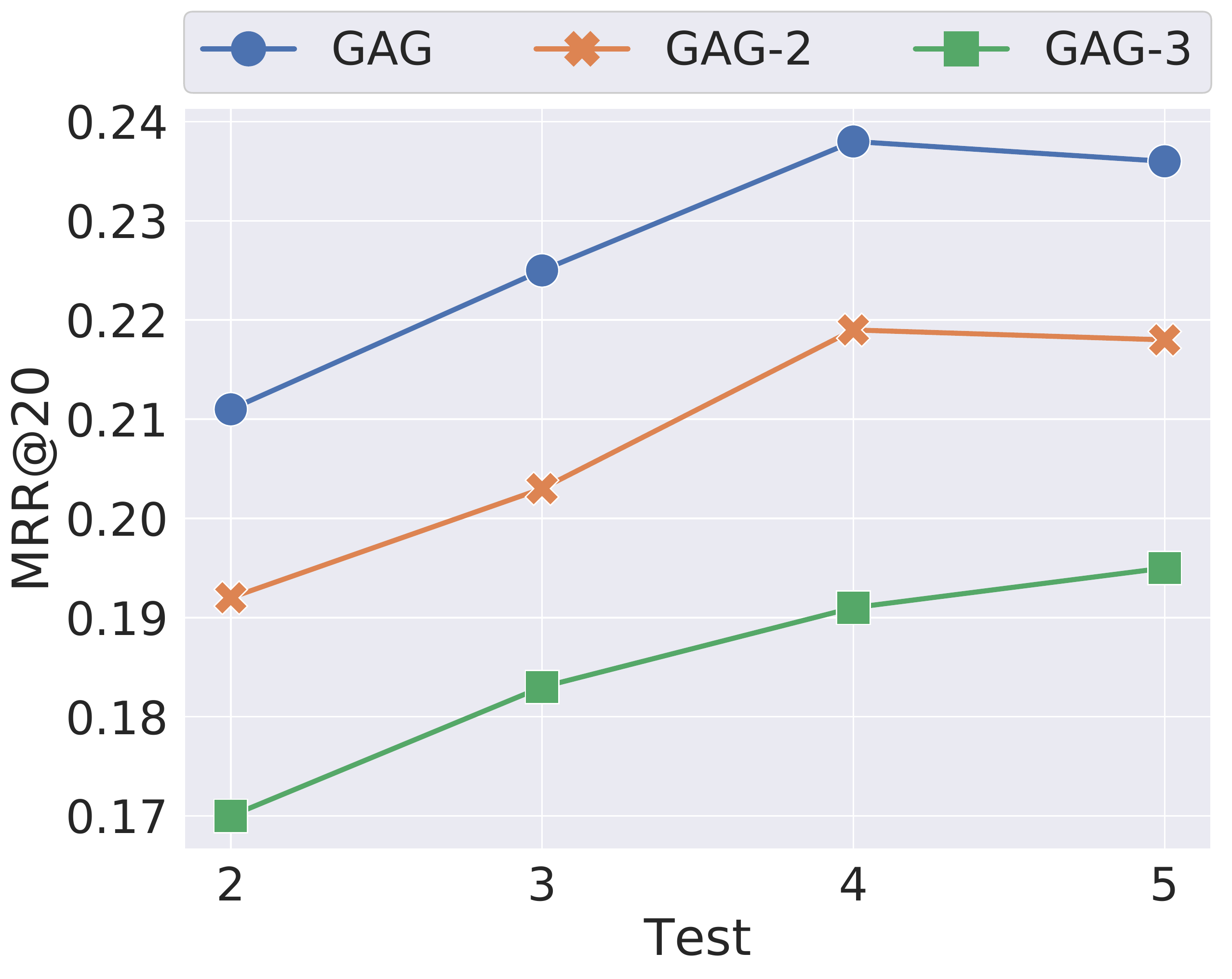}
    }
    \subfigure[Recall@20 on LastFM.]{
    \label{fig:lastfm_num_R20}
    \includegraphics[width=0.46\linewidth]{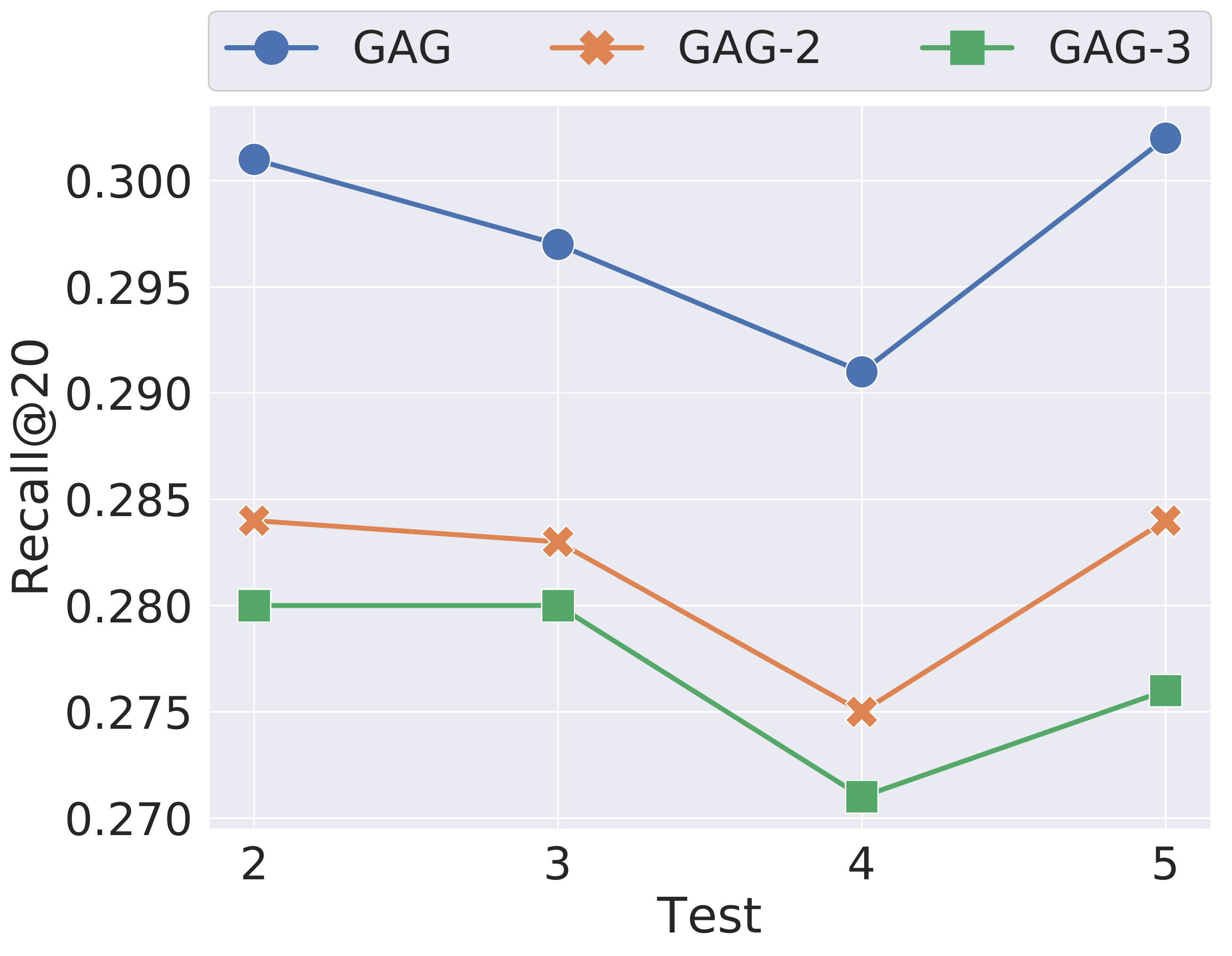}
    }
    \subfigure[MRR@20 on LastFM.]{
    \label{fig:lastfm_num_M20}
    \includegraphics[width=0.47\linewidth]{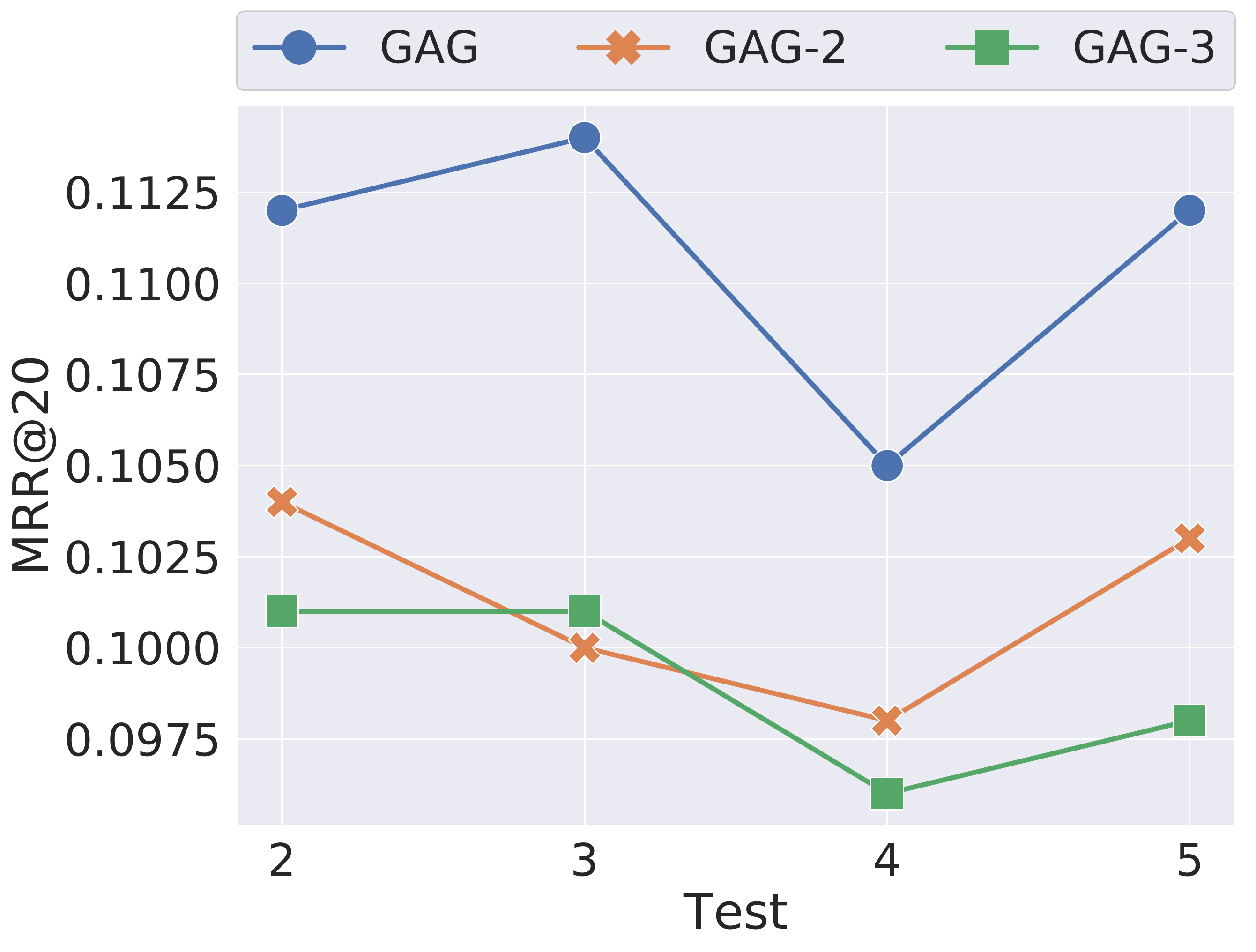}
    }
    \vspace{-0.5cm}
    \caption{Results of different numbers of layers.}
    \label{fig:num}
\end{figure}

In Fig.~\ref{fig:num}, the result of different layers is presented. Generally, GNN models always suffer from an increase in the depth of the model because of the gradient explosion. In our experiment, the performance of the GAG model decreases as the model goes deeper, which is consistent with the common observation. Furthermore, the connectivity of sessions is smaller than the traditional graph data, which also limits the power of deeper GNN models.

\section{Conclusion}
In this paper, we proposed a GAG model with a Wasserstein reservoir to perform SSR. We addressed the problem of how to preserve users' long-term interests by introducing the global attribute and the GAG layer. We designed an effective and generic Wasserstein reservoir, which samples sessions according to the Wasserstein distance between their recommendation results and the real interactions. In the future, it is significant to investigate how to incorporate the cross-session information for the SSR problem.

\section{Acknowledgments}
The work has been supported by Australian Research Council (Grant No. DP190101985, DP170103954 and FT200100825).

\bibliographystyle{ACM-Reference-Format}
\bibliography{sigir.bib}


\begin{thebibliography}{44}


\ifx \showCODEN    \undefined \def \showCODEN     #1{\unskip}     \fi
\ifx \showDOI      \undefined \def \showDOI       #1{#1}\fi
\ifx \showISBNx    \undefined \def \showISBNx     #1{\unskip}     \fi
\ifx \showISBNxiii \undefined \def \showISBNxiii  #1{\unskip}     \fi
\ifx \showISSN     \undefined \def \showISSN      #1{\unskip}     \fi
\ifx \showLCCN     \undefined \def \showLCCN      #1{\unskip}     \fi
\ifx \shownote     \undefined \def \shownote      #1{#1}          \fi
\ifx \showarticletitle \undefined \def \showarticletitle #1{#1}   \fi
\ifx \showURL      \undefined \def \showURL       {\relax}        \fi
\providecommand\bibfield[2]{#2}
\providecommand\bibinfo[2]{#2}
\providecommand\natexlab[1]{#1}
\providecommand\showeprint[2][]{arXiv:#2}

\bibitem[\protect\citeauthoryear{Battaglia, Hamrick, Bapst, Sanchez{-}Gonzalez,
  Zambaldi, Malinowski, Tacchetti, Raposo, Santoro, Faulkner,
  G{\"{u}}l{\c{c}}ehre, Song, Ballard, Gilmer, Dahl, Vaswani, Allen, Nash,
  Langston, Dyer, Heess, Wierstra, Kohli, Botvinick, Vinyals, Li, and
  Pascanu}{Battaglia et~al\mbox{.}}{2018}]%
        {battaglia2018relational}
\bibfield{author}{\bibinfo{person}{Peter~W. Battaglia},
  \bibinfo{person}{Jessica~B. Hamrick}, \bibinfo{person}{Victor Bapst},
  \bibinfo{person}{Alvaro Sanchez{-}Gonzalez},
  \bibinfo{person}{Vin{\'{\i}}cius~Flores Zambaldi}, \bibinfo{person}{Mateusz
  Malinowski}, \bibinfo{person}{Andrea Tacchetti}, \bibinfo{person}{David
  Raposo}, \bibinfo{person}{Adam Santoro}, \bibinfo{person}{Ryan Faulkner},
  \bibinfo{person}{{\c{C}}aglar G{\"{u}}l{\c{c}}ehre}, \bibinfo{person}{Francis
  Song}, \bibinfo{person}{Andrew~J. Ballard}, \bibinfo{person}{Justin Gilmer},
  \bibinfo{person}{George~E. Dahl}, \bibinfo{person}{Ashish Vaswani},
  \bibinfo{person}{Kelsey Allen}, \bibinfo{person}{Charles Nash},
  \bibinfo{person}{Victoria Langston}, \bibinfo{person}{Chris Dyer},
  \bibinfo{person}{Nicolas Heess}, \bibinfo{person}{Daan Wierstra},
  \bibinfo{person}{Pushmeet Kohli}, \bibinfo{person}{Matthew Botvinick},
  \bibinfo{person}{Oriol Vinyals}, \bibinfo{person}{Yujia Li}, {and}
  \bibinfo{person}{Razvan Pascanu}.} \bibinfo{year}{2018}\natexlab{}.
\newblock \showarticletitle{Relational inductive biases, deep learning, and
  graph networks}.
\newblock \bibinfo{journal}{\emph{CoRR}}  \bibinfo{volume}{abs/1806.01261}
  (\bibinfo{year}{2018}).
\newblock


\bibitem[\protect\citeauthoryear{Celma}{Celma}{2010}]%
        {Celma10Music}
\bibfield{author}{\bibinfo{person}{{\`{O}}scar Celma}.}
  \bibinfo{year}{2010}\natexlab{}.
\newblock \bibinfo{booktitle}{\emph{Music Recommendation and Discovery - The
  Long Tail, Long Fail, and Long Play in the Digital Music Space}}.
\newblock \bibinfo{publisher}{Springer}.
\newblock


\bibitem[\protect\citeauthoryear{Chang, Zhang, Tang, Yin, Chang,
  Hasegawa{-}Johnson, and Huang}{Chang et~al\mbox{.}}{2017}]%
        {ChangZTYCHH17}
\bibfield{author}{\bibinfo{person}{Shiyu Chang}, \bibinfo{person}{Yang Zhang},
  \bibinfo{person}{Jiliang Tang}, \bibinfo{person}{Dawei Yin},
  \bibinfo{person}{Yi Chang}, \bibinfo{person}{Mark~A. Hasegawa{-}Johnson},
  {and} \bibinfo{person}{Thomas~S. Huang}.} \bibinfo{year}{2017}\natexlab{}.
\newblock \showarticletitle{Streaming Recommender Systems}. In
  \bibinfo{booktitle}{\emph{WWW}}.
\newblock


\bibitem[\protect\citeauthoryear{Chen, Yin, Yao, and Cui}{Chen
  et~al\mbox{.}}{2013}]%
        {ChenYYC13}
\bibfield{author}{\bibinfo{person}{Chen Chen}, \bibinfo{person}{Hongzhi Yin},
  \bibinfo{person}{Junjie Yao}, {and} \bibinfo{person}{Bin Cui}.}
  \bibinfo{year}{2013}\natexlab{}.
\newblock \showarticletitle{TeRec: {A} Temporal Recommender System Over Tweet
  Stream}.
\newblock \bibinfo{journal}{\emph{{PVLDB}}} \bibinfo{volume}{6},
  \bibinfo{number}{12} (\bibinfo{year}{2013}).
\newblock


\bibitem[\protect\citeauthoryear{Chung, Gulcehre, Cho, and Bengio}{Chung
  et~al\mbox{.}}{2014}]%
        {chung2014empirical}
\bibfield{author}{\bibinfo{person}{Junyoung Chung}, \bibinfo{person}{Caglar
  Gulcehre}, \bibinfo{person}{Kyunghyun Cho}, {and} \bibinfo{person}{Yoshua
  Bengio}.} \bibinfo{year}{2014}\natexlab{}.
\newblock \showarticletitle{Empirical evaluation of gated recurrent neural
  networks on sequence modeling}. In \bibinfo{booktitle}{\emph{NIPS}}.
\newblock


\bibitem[\protect\citeauthoryear{Diaz{-}Aviles, Drumond, Schmidt{-}Thieme, and
  Nejdl}{Diaz{-}Aviles et~al\mbox{.}}{2012}]%
        {Diaz-AvilesDSN12}
\bibfield{author}{\bibinfo{person}{Ernesto Diaz{-}Aviles},
  \bibinfo{person}{Lucas Drumond}, \bibinfo{person}{Lars Schmidt{-}Thieme},
  {and} \bibinfo{person}{Wolfgang Nejdl}.} \bibinfo{year}{2012}\natexlab{}.
\newblock \showarticletitle{Real-time top-n recommendation in social streams}.
  In \bibinfo{booktitle}{\emph{RecSys}}.
\newblock


\bibitem[\protect\citeauthoryear{Gilmer, Schoenholz, Riley, Vinyals, and
  Dahl}{Gilmer et~al\mbox{.}}{2017}]%
        {gilmer2017neural}
\bibfield{author}{\bibinfo{person}{Justin Gilmer}, \bibinfo{person}{Samuel~S.
  Schoenholz}, \bibinfo{person}{Patrick~F. Riley}, \bibinfo{person}{Oriol
  Vinyals}, {and} \bibinfo{person}{George~E. Dahl}.}
  \bibinfo{year}{2017}\natexlab{}.
\newblock \showarticletitle{Neural Message Passing for Quantum Chemistry}. In
  \bibinfo{booktitle}{\emph{ICML}}.
\newblock


\bibitem[\protect\citeauthoryear{Gori, Monfardini, and Scarselli}{Gori
  et~al\mbox{.}}{2005}]%
        {gori2005new}
\bibfield{author}{\bibinfo{person}{Marco Gori}, \bibinfo{person}{Gabriele
  Monfardini}, {and} \bibinfo{person}{Franco Scarselli}.}
  \bibinfo{year}{2005}\natexlab{}.
\newblock \showarticletitle{A new model for learning in graph domains}. In
  \bibinfo{booktitle}{\emph{IJCNN}}, Vol.~\bibinfo{volume}{2}.
\newblock


\bibitem[\protect\citeauthoryear{Guo, Yin, Wang, Chen, Zhou, and Hung}{Guo
  et~al\mbox{.}}{2019}]%
        {guo2019streaming}
\bibfield{author}{\bibinfo{person}{Lei Guo}, \bibinfo{person}{Hongzhi Yin},
  \bibinfo{person}{Qinyong Wang}, \bibinfo{person}{Tong Chen},
  \bibinfo{person}{Alexander Zhou}, {and} \bibinfo{person}{Nguyen Quoc~Viet
  Hung}.} \bibinfo{year}{2019}\natexlab{}.
\newblock \showarticletitle{Streaming Session-based Recommendation}. In
  \bibinfo{booktitle}{\emph{SIGKDD}}.
\newblock


\bibitem[\protect\citeauthoryear{Hamilton, Ying, and Leskovec}{Hamilton
  et~al\mbox{.}}{2017}]%
        {hamilton2017inductive}
\bibfield{author}{\bibinfo{person}{William~L. Hamilton},
  \bibinfo{person}{Zhitao Ying}, {and} \bibinfo{person}{Jure Leskovec}.}
  \bibinfo{year}{2017}\natexlab{}.
\newblock \showarticletitle{Inductive Representation Learning on Large Graphs}.
  In \bibinfo{booktitle}{\emph{NIPS}}.
\newblock


\bibitem[\protect\citeauthoryear{Hamrick, Allen, Bapst, Zhu, McKee, Tenenbaum,
  and Battaglia}{Hamrick et~al\mbox{.}}{2018}]%
        {HamrickABZMTB18}
\bibfield{author}{\bibinfo{person}{Jessica~B. Hamrick}, \bibinfo{person}{Kelsey
  Allen}, \bibinfo{person}{Victor Bapst}, \bibinfo{person}{Tina Zhu},
  \bibinfo{person}{Kevin~R. McKee}, \bibinfo{person}{Josh Tenenbaum}, {and}
  \bibinfo{person}{Peter Battaglia}.} \bibinfo{year}{2018}\natexlab{}.
\newblock \showarticletitle{Relational inductive bias for physical construction
  in humans and machines}. In \bibinfo{booktitle}{\emph{CogSci}}.
\newblock


\bibitem[\protect\citeauthoryear{He, Zhang, Kan, and Chua}{He
  et~al\mbox{.}}{2016}]%
        {HeZKC16}
\bibfield{author}{\bibinfo{person}{Xiangnan He}, \bibinfo{person}{Hanwang
  Zhang}, \bibinfo{person}{Min{-}Yen Kan}, {and} \bibinfo{person}{Tat{-}Seng
  Chua}.} \bibinfo{year}{2016}\natexlab{}.
\newblock \showarticletitle{Fast Matrix Factorization for Online Recommendation
  with Implicit Feedback}. In \bibinfo{booktitle}{\emph{SIGIR}}.
\newblock


\bibitem[\protect\citeauthoryear{Hidasi, Karatzoglou, Baltrunas, and
  Tikk}{Hidasi et~al\mbox{.}}{2016}]%
        {hidasi2015session}
\bibfield{author}{\bibinfo{person}{Bal{\'{a}}zs Hidasi},
  \bibinfo{person}{Alexandros Karatzoglou}, \bibinfo{person}{Linas Baltrunas},
  {and} \bibinfo{person}{Domonkos Tikk}.} \bibinfo{year}{2016}\natexlab{}.
\newblock \showarticletitle{Session-based Recommendations with Recurrent Neural
  Networks}. In \bibinfo{booktitle}{\emph{ICLR}}.
\newblock


\bibitem[\protect\citeauthoryear{Jugovac, Jannach, and Karimi}{Jugovac
  et~al\mbox{.}}{2018}]%
        {JugovacJK18}
\bibfield{author}{\bibinfo{person}{Michael Jugovac}, \bibinfo{person}{Dietmar
  Jannach}, {and} \bibinfo{person}{Mozhgan Karimi}.}
  \bibinfo{year}{2018}\natexlab{}.
\newblock \showarticletitle{Streamingrec: a framework for benchmarking
  stream-based news recommenders}. In \bibinfo{booktitle}{\emph{RecSys}}.
\newblock


\bibitem[\protect\citeauthoryear{Kang and McAuley}{Kang and McAuley}{2018}]%
        {KangM18}
\bibfield{author}{\bibinfo{person}{Wang{-}Cheng Kang} {and}
  \bibinfo{person}{Julian~J. McAuley}.} \bibinfo{year}{2018}\natexlab{}.
\newblock \showarticletitle{Self-Attentive Sequential Recommendation}. In
  \bibinfo{booktitle}{\emph{ICDM}}.
\newblock


\bibitem[\protect\citeauthoryear{Kingma and Ba}{Kingma and Ba}{2015}]%
        {adam}
\bibfield{author}{\bibinfo{person}{Diederik~P. Kingma} {and}
  \bibinfo{person}{Jimmy Ba}.} \bibinfo{year}{2015}\natexlab{}.
\newblock \showarticletitle{Adam: {A} Method for Stochastic Optimization}. In
  \bibinfo{booktitle}{\emph{ICLR}}.
\newblock


\bibitem[\protect\citeauthoryear{Kipf and Welling}{Kipf and Welling}{2017}]%
        {kipf2017semi}
\bibfield{author}{\bibinfo{person}{Thomas~N. Kipf} {and} \bibinfo{person}{Max
  Welling}.} \bibinfo{year}{2017}\natexlab{}.
\newblock \showarticletitle{Semi-Supervised Classification with Graph
  Convolutional Networks}. In \bibinfo{booktitle}{\emph{ICLR}}.
\newblock


\bibitem[\protect\citeauthoryear{Kullback and Leibler}{Kullback and
  Leibler}{1951}]%
        {kullback1951}
\bibfield{author}{\bibinfo{person}{S. Kullback} {and} \bibinfo{person}{R.~A.
  Leibler}.} \bibinfo{year}{1951}\natexlab{}.
\newblock \showarticletitle{On Information and Sufficiency}.
\newblock \bibinfo{journal}{\emph{Ann. Math. Statist.}} \bibinfo{volume}{22},
  \bibinfo{number}{1} (\bibinfo{year}{1951}), \bibinfo{pages}{79--86}.
\newblock


\bibitem[\protect\citeauthoryear{Li, Ren, Chen, Ren, Lian, and Ma}{Li
  et~al\mbox{.}}{2017}]%
        {li2017neural}
\bibfield{author}{\bibinfo{person}{Jing Li}, \bibinfo{person}{Pengjie Ren},
  \bibinfo{person}{Zhumin Chen}, \bibinfo{person}{Zhaochun Ren},
  \bibinfo{person}{Tao Lian}, {and} \bibinfo{person}{Jun Ma}.}
  \bibinfo{year}{2017}\natexlab{}.
\newblock \showarticletitle{Neural Attentive Session-based Recommendation}. In
  \bibinfo{booktitle}{\emph{CIKM}}.
\newblock


\bibitem[\protect\citeauthoryear{Liu, Zeng, Mokhosi, and Zhang}{Liu
  et~al\mbox{.}}{2018}]%
        {Liu18STAMP}
\bibfield{author}{\bibinfo{person}{Qiao Liu}, \bibinfo{person}{Yifu Zeng},
  \bibinfo{person}{Refuoe Mokhosi}, {and} \bibinfo{person}{Haibin Zhang}.}
  \bibinfo{year}{2018}\natexlab{}.
\newblock \showarticletitle{{STAMP:} Short-Term Attention/Memory Priority Model
  for Session-based Recommendation}. In \bibinfo{booktitle}{\emph{SIGKDD}}.
\newblock


\bibitem[\protect\citeauthoryear{Ma, Kang, and Liu}{Ma et~al\mbox{.}}{2019}]%
        {MaKL19}
\bibfield{author}{\bibinfo{person}{Chen Ma}, \bibinfo{person}{Peng Kang}, {and}
  \bibinfo{person}{Xue Liu}.} \bibinfo{year}{2019}\natexlab{}.
\newblock \showarticletitle{Hierarchical Gating Networks for Sequential
  Recommendation}. In \bibinfo{booktitle}{\emph{SIGKDD}}.
\newblock


\bibitem[\protect\citeauthoryear{Qiu, Li, Huang, and Yin}{Qiu
  et~al\mbox{.}}{2019}]%
        {qiu2019rethinking}
\bibfield{author}{\bibinfo{person}{Ruihong Qiu}, \bibinfo{person}{Jingjing Li},
  \bibinfo{person}{Zi Huang}, {and} \bibinfo{person}{Hongzhi Yin}.}
  \bibinfo{year}{2019}\natexlab{}.
\newblock \showarticletitle{Rethinking the Item Order in Session-based
  Recommendation with Graph Neural Networks}. In
  \bibinfo{booktitle}{\emph{CIKM}}.
\newblock


\bibitem[\protect\citeauthoryear{Qiu, Li, Huang, and Yin}{Qiu
  et~al\mbox{.}}{2020}]%
        {QiuTOIS20}
\bibfield{author}{\bibinfo{person}{Ruihong Qiu}, \bibinfo{person}{Jingjing Li},
  \bibinfo{person}{Zi Huang}, {and} \bibinfo{person}{Hongzhi Yin}.}
  \bibinfo{year}{2020}\natexlab{}.
\newblock \showarticletitle{Exploiting Cross-Session Information for
  Session-based Recommendation with Graph Neural Networks}.
\newblock \bibinfo{journal}{\emph{{ACM} Trans. Inf. Syst.}}
  (\bibinfo{year}{2020}).
\newblock


\bibitem[\protect\citeauthoryear{Rendle, Freudenthaler, Gantner, and
  Schmidt{-}Thieme}{Rendle et~al\mbox{.}}{2009}]%
        {Rendle09Bayesian}
\bibfield{author}{\bibinfo{person}{Steffen Rendle}, \bibinfo{person}{Christoph
  Freudenthaler}, \bibinfo{person}{Zeno Gantner}, {and} \bibinfo{person}{Lars
  Schmidt{-}Thieme}.} \bibinfo{year}{2009}\natexlab{}.
\newblock \showarticletitle{{BPR:} Bayesian Personalized Ranking from Implicit
  Feedback}. In \bibinfo{booktitle}{\emph{UAI}}.
\newblock


\bibitem[\protect\citeauthoryear{Rendle and Schmidt{-}Thieme}{Rendle and
  Schmidt{-}Thieme}{2008}]%
        {RendleS08}
\bibfield{author}{\bibinfo{person}{Steffen Rendle} {and} \bibinfo{person}{Lars
  Schmidt{-}Thieme}.} \bibinfo{year}{2008}\natexlab{}.
\newblock \showarticletitle{Online-updating regularized kernel matrix
  factorization models for large-scale recommender systems}. In
  \bibinfo{booktitle}{\emph{RecSys}}.
\newblock


\bibitem[\protect\citeauthoryear{Rubner, Tomasi, and Guibas}{Rubner
  et~al\mbox{.}}{2000}]%
        {RubnerTG00}
\bibfield{author}{\bibinfo{person}{Yossi Rubner}, \bibinfo{person}{Carlo
  Tomasi}, {and} \bibinfo{person}{Leonidas~J. Guibas}.}
  \bibinfo{year}{2000}\natexlab{}.
\newblock \showarticletitle{The Earth Mover's Distance as a Metric for Image
  Retrieval}.
\newblock \bibinfo{journal}{\emph{International Journal of Computer Vision}}
  \bibinfo{volume}{40}, \bibinfo{number}{2} (\bibinfo{year}{2000}),
  \bibinfo{pages}{99--121}.
\newblock


\bibitem[\protect\citeauthoryear{Sanchez{-}Gonzalez, Heess, Springenberg,
  Merel, Riedmiller, Hadsell, and Battaglia}{Sanchez{-}Gonzalez
  et~al\mbox{.}}{2018}]%
        {Sanchez-Gonzalez18}
\bibfield{author}{\bibinfo{person}{Alvaro Sanchez{-}Gonzalez},
  \bibinfo{person}{Nicolas Heess}, \bibinfo{person}{Jost~Tobias Springenberg},
  \bibinfo{person}{Josh Merel}, \bibinfo{person}{Martin~A. Riedmiller},
  \bibinfo{person}{Raia Hadsell}, {and} \bibinfo{person}{Peter Battaglia}.}
  \bibinfo{year}{2018}\natexlab{}.
\newblock \showarticletitle{Graph Networks as Learnable Physics Engines for
  Inference and Control}. In \bibinfo{booktitle}{\emph{ICML}}.
\newblock


\bibitem[\protect\citeauthoryear{Scarselli, Gori, Tsoi, Hagenbuchner, and
  Monfardini}{Scarselli et~al\mbox{.}}{2009}]%
        {scarselli2009graph}
\bibfield{author}{\bibinfo{person}{Franco Scarselli}, \bibinfo{person}{Marco
  Gori}, \bibinfo{person}{Ah~Chung Tsoi}, \bibinfo{person}{Markus
  Hagenbuchner}, {and} \bibinfo{person}{Gabriele Monfardini}.}
  \bibinfo{year}{2009}\natexlab{}.
\newblock \showarticletitle{The Graph Neural Network Model}.
\newblock \bibinfo{journal}{\emph{{IEEE} Trans. Neural Networks}}
  \bibinfo{volume}{20}, \bibinfo{number}{1} (\bibinfo{year}{2009}).
\newblock


\bibitem[\protect\citeauthoryear{Shani, Brafman, and Heckerman}{Shani
  et~al\mbox{.}}{2002}]%
        {shani2005mdp}
\bibfield{author}{\bibinfo{person}{Guy Shani}, \bibinfo{person}{Ronen~I.
  Brafman}, {and} \bibinfo{person}{David Heckerman}.}
  \bibinfo{year}{2002}\natexlab{}.
\newblock \showarticletitle{An MDP-based Recommender System}. In
  \bibinfo{booktitle}{\emph{UAI}}.
\newblock


\bibitem[\protect\citeauthoryear{Sun, Liu, Wu, Pei, Lin, Ou, and Jiang}{Sun
  et~al\mbox{.}}{2019a}]%
        {SunLWPLOJ19}
\bibfield{author}{\bibinfo{person}{Fei Sun}, \bibinfo{person}{Jun Liu},
  \bibinfo{person}{Jian Wu}, \bibinfo{person}{Changhua Pei},
  \bibinfo{person}{Xiao Lin}, \bibinfo{person}{Wenwu Ou}, {and}
  \bibinfo{person}{Peng Jiang}.} \bibinfo{year}{2019}\natexlab{a}.
\newblock \showarticletitle{BERT4Rec: Sequential Recommendation with
  Bidirectional Encoder Representations from Transformer}. In
  \bibinfo{booktitle}{\emph{CIKM}}.
\newblock


\bibitem[\protect\citeauthoryear{Sun, Zhang, Ma, Coates, Guo, Tang, and He}{Sun
  et~al\mbox{.}}{2019b}]%
        {SunZMCGTH19}
\bibfield{author}{\bibinfo{person}{Jianing Sun}, \bibinfo{person}{Yingxue
  Zhang}, \bibinfo{person}{Chen Ma}, \bibinfo{person}{Mark Coates},
  \bibinfo{person}{Huifeng Guo}, \bibinfo{person}{Ruiming Tang}, {and}
  \bibinfo{person}{Xiuqiang He}.} \bibinfo{year}{2019}\natexlab{b}.
\newblock \showarticletitle{Multi-graph Convolution Collaborative Filtering}.
  In \bibinfo{booktitle}{\emph{ICDM}}.
\newblock


\bibitem[\protect\citeauthoryear{Tang and Wang}{Tang and Wang}{[n.d.]}]%
        {TangW18}
\bibfield{author}{\bibinfo{person}{Jiaxi Tang} {and} \bibinfo{person}{Ke
  Wang}.} \bibinfo{year}{[n.d.]}\natexlab{}.
\newblock \showarticletitle{Personalized Top-N Sequential Recommendation via
  Convolutional Sequence Embedding}.
\newblock


\bibitem[\protect\citeauthoryear{Vitter}{Vitter}{1985}]%
        {Vitter85Random}
\bibfield{author}{\bibinfo{person}{Jeffrey~Scott Vitter}.}
  \bibinfo{year}{1985}\natexlab{}.
\newblock \showarticletitle{Random Sampling with a Reservoir}.
\newblock \bibinfo{journal}{\emph{{ACM} Trans. Math. Softw.}}
  \bibinfo{volume}{11}, \bibinfo{number}{1} (\bibinfo{year}{1985}),
  \bibinfo{pages}{37--57}.
\newblock


\bibitem[\protect\citeauthoryear{Wang, Yin, Hu, Lian, Wang, and Huang}{Wang
  et~al\mbox{.}}{2018a}]%
        {WangYHLWH18}
\bibfield{author}{\bibinfo{person}{Qinyong Wang}, \bibinfo{person}{Hongzhi
  Yin}, \bibinfo{person}{Zhiting Hu}, \bibinfo{person}{Defu Lian},
  \bibinfo{person}{Hao Wang}, {and} \bibinfo{person}{Zi Huang}.}
  \bibinfo{year}{2018}\natexlab{a}.
\newblock \showarticletitle{Neural Memory Streaming Recommender Networks with
  Adversarial Training}. In \bibinfo{booktitle}{\emph{SIGKDD}}.
\newblock


\bibitem[\protect\citeauthoryear{Wang, Yin, Huang, Wang, Du, and Nguyen}{Wang
  et~al\mbox{.}}{2018b}]%
        {WangYHWDN18}
\bibfield{author}{\bibinfo{person}{Weiqing Wang}, \bibinfo{person}{Hongzhi
  Yin}, \bibinfo{person}{Zi Huang}, \bibinfo{person}{Qinyong Wang},
  \bibinfo{person}{Xingzhong Du}, {and} \bibinfo{person}{Quoc Viet~Hung
  Nguyen}.} \bibinfo{year}{2018}\natexlab{b}.
\newblock \showarticletitle{Streaming Ranking Based Recommender Systems}. In
  \bibinfo{booktitle}{\emph{SIGIR}}.
\newblock


\bibitem[\protect\citeauthoryear{Wang, Yin, Sadiq, Chen, Xie, and Zhou}{Wang
  et~al\mbox{.}}{2016}]%
        {WangYSCXZ16}
\bibfield{author}{\bibinfo{person}{Weiqing Wang}, \bibinfo{person}{Hongzhi
  Yin}, \bibinfo{person}{Shazia~Wasim Sadiq}, \bibinfo{person}{Ling Chen},
  \bibinfo{person}{Min Xie}, {and} \bibinfo{person}{Xiaofang Zhou}.}
  \bibinfo{year}{2016}\natexlab{}.
\newblock \showarticletitle{SPORE: {A} Sequential Personalized Spatial Item
  Recommender System}. In \bibinfo{booktitle}{\emph{ICDE}}.
\newblock


\bibitem[\protect\citeauthoryear{Wang, He, Wang, Feng, and Chua}{Wang
  et~al\mbox{.}}{2019}]%
        {Wang0WFC19}
\bibfield{author}{\bibinfo{person}{Xiang Wang}, \bibinfo{person}{Xiangnan He},
  \bibinfo{person}{Meng Wang}, \bibinfo{person}{Fuli Feng}, {and}
  \bibinfo{person}{Tat{-}Seng Chua}.} \bibinfo{year}{2019}\natexlab{}.
\newblock \showarticletitle{Neural Graph Collaborative Filtering}. In
  \bibinfo{booktitle}{\emph{SIGIR}}.
\newblock


\bibitem[\protect\citeauthoryear{Wu, Tang, Zhu, Wang, Xie, and Tan}{Wu
  et~al\mbox{.}}{2019}]%
        {wu2018session}
\bibfield{author}{\bibinfo{person}{Shu Wu}, \bibinfo{person}{Yuyuan Tang},
  \bibinfo{person}{Yanqiao Zhu}, \bibinfo{person}{Liang Wang},
  \bibinfo{person}{Xing Xie}, {and} \bibinfo{person}{Tieniu Tan}.}
  \bibinfo{year}{2019}\natexlab{}.
\newblock \showarticletitle{Session-based Recommendation with Graph Neural
  Networks}. In \bibinfo{booktitle}{\emph{AAAI}}.
\newblock


\bibitem[\protect\citeauthoryear{Xu, Zhao, Liu, Sheng, Xu, Zhuang, Fang, and
  Zhou}{Xu et~al\mbox{.}}{2019}]%
        {XuZLSXZFZ19}
\bibfield{author}{\bibinfo{person}{Chengfeng Xu}, \bibinfo{person}{Pengpeng
  Zhao}, \bibinfo{person}{Yanchi Liu}, \bibinfo{person}{Victor~S. Sheng},
  \bibinfo{person}{Jiajie Xu}, \bibinfo{person}{Fuzhen Zhuang},
  \bibinfo{person}{Junhua Fang}, {and} \bibinfo{person}{Xiaofang Zhou}.}
  \bibinfo{year}{2019}\natexlab{}.
\newblock \showarticletitle{Graph Contextualized Self-Attention Network for
  Session-based Recommendation}. In \bibinfo{booktitle}{\emph{IJCAI}}.
\newblock


\bibitem[\protect\citeauthoryear{Yin, Cui, Chen, Hu, and Zhou}{Yin
  et~al\mbox{.}}{2015}]%
        {YinCCHZ15}
\bibfield{author}{\bibinfo{person}{Hongzhi Yin}, \bibinfo{person}{Bin Cui},
  \bibinfo{person}{Ling Chen}, \bibinfo{person}{Zhiting Hu}, {and}
  \bibinfo{person}{Xiaofang Zhou}.} \bibinfo{year}{2015}\natexlab{}.
\newblock \showarticletitle{Dynamic User Modeling in Social Media Systems}.
\newblock \bibinfo{journal}{\emph{{ACM} Trans. Inf. Syst.}}
  \bibinfo{volume}{33}, \bibinfo{number}{3} (\bibinfo{year}{2015}).
\newblock


\bibitem[\protect\citeauthoryear{Yin, Cui, Zhou, Wang, Huang, and Sadiq}{Yin
  et~al\mbox{.}}{2016}]%
        {YinCZWHS16}
\bibfield{author}{\bibinfo{person}{Hongzhi Yin}, \bibinfo{person}{Bin Cui},
  \bibinfo{person}{Xiaofang Zhou}, \bibinfo{person}{Weiqing Wang},
  \bibinfo{person}{Zi Huang}, {and} \bibinfo{person}{Shazia~W. Sadiq}.}
  \bibinfo{year}{2016}\natexlab{}.
\newblock \showarticletitle{Joint Modeling of User Check-in Behaviors for
  Real-time Point-of-Interest Recommendation}.
\newblock \bibinfo{journal}{\emph{{ACM} Trans. Inf. Syst.}}
  \bibinfo{volume}{35}, \bibinfo{number}{2} (\bibinfo{year}{2016}).
\newblock


\bibitem[\protect\citeauthoryear{Ying, He, Chen, Eksombatchai, Hamilton, and
  Leskovec}{Ying et~al\mbox{.}}{2018}]%
        {YingHCEHL18}
\bibfield{author}{\bibinfo{person}{Rex Ying}, \bibinfo{person}{Ruining He},
  \bibinfo{person}{Kaifeng Chen}, \bibinfo{person}{Pong Eksombatchai},
  \bibinfo{person}{William~L. Hamilton}, {and} \bibinfo{person}{Jure
  Leskovec}.} \bibinfo{year}{2018}\natexlab{}.
\newblock \showarticletitle{Graph Convolutional Neural Networks for Web-Scale
  Recommender Systems}. In \bibinfo{booktitle}{\emph{SIGKDD}}.
\newblock


\bibitem[\protect\citeauthoryear{Zhang, Jiang, Wang, Feng, and Xie}{Zhang
  et~al\mbox{.}}{2019}]%
        {Zhang00FX19}
\bibfield{author}{\bibinfo{person}{Lingling Zhang}, \bibinfo{person}{Hong
  Jiang}, \bibinfo{person}{Fang Wang}, \bibinfo{person}{Dan Feng}, {and}
  \bibinfo{person}{Yanwen Xie}.} \bibinfo{year}{2019}\natexlab{}.
\newblock \showarticletitle{T-Sample: {A} Dual Reservoir-Based Sampling Method
  for Characterizing Large Graph Streams}. In \bibinfo{booktitle}{\emph{ICDE}}.
\newblock


\bibitem[\protect\citeauthoryear{Zimdars, Chickering, and Meek}{Zimdars
  et~al\mbox{.}}{2001}]%
        {zimdars2001using}
\bibfield{author}{\bibinfo{person}{Andrew Zimdars},
  \bibinfo{person}{David~Maxwell Chickering}, {and}
  \bibinfo{person}{Christopher Meek}.} \bibinfo{year}{2001}\natexlab{}.
\newblock \showarticletitle{Using Temporal Data for Making Recommendations}. In
  \bibinfo{booktitle}{\emph{UAI}}.
\newblock


\end{thebibliography}

\end{document}